\documentclass[aps,prl,twocolumn,showpacs,superscriptaddress,groupedaddress]{revtex4-1}

\usepackage{amsfonts}
\usepackage{amsmath}
\usepackage{amssymb}
\usepackage{epsfig}
\usepackage{graphicx}
\usepackage{colordvi}
\usepackage{graphics}
\usepackage{color}

\usepackage{subfig}

\newcommand{\bff}{{\bf f}}%
\newcommand{\bfu}{{\bf u}}%
\newcommand{\bfv}{{\bf v}}%
\newcommand{\bfx}{{\bf x}}%
\newcommand{\bfomega}{\boldsymbol{\omega}}%
\newfont{\tenbfit}{cmbx10}%

\newfont{\tenbbb}{msbm10}%
\newfont{\svnbbb}{msbm8}%
\newcommand{\half}{{\textstyle{\frac{1}{2}}}}

\newcommand{\Curl}{\hbox{\rm curl}\mskip2mu}

\newcommand{\lj}{\mbox{$[\kern-0.1478125em[$}}
\newcommand{\rj}{\mbox{$]\kern-0.1478125em]$}}
\newcommand{\la}{\mbox{$\langle\kern-0.2325em\langle$}}
\newcommand{\ra}{\mbox{$\rangle\kern-0.2325em\rangle$}}
\newcommand{\Blj}{\mbox{$\Big[\kern-0.275em\Big[$}}
\newcommand{\Brj}{\mbox{$\Big]\kern-0.275em\Big]$}}
\newcommand{\Bla}{\mbox{$\Big\langle\kern-0.425em\Big\langle$}}
\newcommand{\Bra}{\mbox{$\Big\rangle\kern-0.425em\Big\rangle$}}

\newcommand{\captionfonts}{\footnotesize}

\makeatletter  
\long\def\@makecaption#1#2{%
  \vskip\abovecaptionskip
  \sbox\@tempboxa{{\captionfonts #1: #2}}%
  \ifdim \wd\@tempboxa >\hsize
    {\captionfonts #1: #2\par}
  \else
    \hbox to\hsize{\hfil\box\@tempboxa\hfil}%
  \fi
  \vskip\belowcaptionskip}
\makeatother   

\begin{document}



\title{Motility versus fluctuations in mixtures of self-motile and passive agents}


\author{Denis F.~Hinz$^1$, 
Alexander Panchenko$^2$,
Tae-Yeon Kim$^3$, and Eliot Fried$^1$}

\affiliation{$^1$Mathematical Soft Matter Unit, Okinawa Institute of Science and Technology, Okinawa, Japan 904-0495, $^2$Department of Mathematics, Washington State University, Pullman, WA, 99164, USA, $^3$Department of Mechanical Engineering, University of Washington,  Seattle, WA 98195, USA}


\date{\today}

\begin{abstract}
Many biological systems consist of self-motile and passive agents both of which contribute to overall functionality. However, there are very few studies of the properties of such mixtures. Here we formulate a model for mixtures of self-motile and passive agents and show that the model gives rise to three different dynamical phases: a disordered mesoturbulent phase, a polar flocking phase, and a vortical phase characterized by large-scale counterrotating vortices. We use numerical simulations to construct a phase diagram and compare the statistical properties of the different phases of the model with self-motile bacterial suspensions. Our findings afford specific insights regarding the interaction of microorganisms and passive particles and provide novel strategic guidance for efficient technological realizations of artificial active matter.
\end{abstract}




\maketitle

The self-motility of individual agents leads to remarkable features of bacterial suspensions, \emph{in-vitro} networks of protein filaments, and the cytoskeletons of living cells. Likewise, macroscopic active systems, such as animal colonies exhibit swarming, herding, and flocking behaviors that appear to share phenomenological similarities with their microscopic counterparts. Such phenomena include polar ordering, large-scale correlated motion, and intriguing rheological properties~\cite{Marchetti2013}. However, biological systems often consist of multiple species which differ in their motilities and other attributes. For example, the emergence of different phenotypes in microbial biofilms generates heterogeneous populations of bacteria~\cite{VanDitmarsch2013, Monds2009, Hibbing2010, Nadell2010, Xavier2011}. In biological systems such as biofilms individual organisms die, malfunction, or lose their flagella, thereby becoming partially or completely immotile. Despite the ubiquity of systems with heterogeneous motility properties, such mixtures have received little attention. Apart from the work of McCandlish et al.~\cite{McCandlish2012}, who report spontaneous segregation in simulations of self-motile and passive rod-shaped agents, we are unaware of any studies of the properties of mixtures of self-motile and passive agents.

Yet, insights regarding the salient biological and mechanical interactions are of great relevance to understanding biological systems and might enable progress in potential technological applications including, in particular, the design of artificial active matter systems. For example, techniques of synthetic biology and systems biology~\cite{Fu2012, Liu2011, Stricker2008, McDaniel2005, You2004, Kobayashi2004} have made it possible to \emph{fabricate} bacterial strains with engineered gene-regulation circuits that produce predefined spatial and temporal patterns. Similarly, artificial self-motile agents can be realized through catalytically driven Janus particles~\cite{Paxton2005, Mano2005, Gibbs2009, Jiang2010, Volpe2011}. From a technological perspective, it is of key importance to know whether it is possible to use a small number of these potentially difficult to manufacture agents to drive other passive agents and thereby generate desirable flow patterns. Having an understanding of how many active agents are required for such a principle seems particularly crucial, as does knowing how such a principle might be realized most efficiently.

In the present letter we study the criteria for which different dynamical phases may be observed in dense mixtures of self-motile and passive spherical soft-core agents. The motion of an agent $i$ with constant mass $m^{i}$ in a system of $N$ agents is governed by Newton's equations including the interaction force $\bff^{ij} = - \bff^{ji}$ between agents $i$ and $j$, and the external force $\bff^{i}_{e}$ exerted on agent $i$. For $\bff^{ij}$, the short-range interaction forces of dissipative particle dynamics (DPD) are used. Working with reduced units and using the standard parameter values~\cite{Groot1997} for a passive DPD fluid, we set $k_BT = 1.0$, $r_c = 1.0$, $A = 25.0$, $\gamma = 4.5$, and $\rho_{\rm{2D}} = \frac{N}{L^2} = 2.5$ (two-dimensional (2D) analog of $\rho_{\rm{3D}}=4.0$), where $L$ is the dimensionless edge length of the square computational domain. Apart from standard DPD forces, we incorporate self-propulsion through a flocking term
\begin{equation}\label{eq:flocking01}
\bff^{i}_{\rm{F}}= (\alpha -\beta|\bfv^i|^2)\bfv^i,
\end{equation}
 with $\alpha \ge 0 $ the constant self-propulsion force parameter and $\beta \ge 0$ the constant Rayleigh friction parameter \cite{Levine2000, DOrsogna2006, Chuang2007, Chuang2007a, Carrillo2009, Carrillo2010}. To model the features of a mixture, we apply $\bff^{i}_{\rm{F}}$ only on the fraction $\phi$ of active agents and the random contribution of the DPD interactions $\bff^{ij}_{\text{R}}$ only between pairs of the fraction $1-\phi$ of passive agents. We perform simulations with LAMMPS~\cite{Plimpton1995, LAMMPS2013} using periodic boundary conditions and the standard velocity-Verlet~\cite{Verlet1967} time integration scheme with a dimenesionless integration timestep of $\Delta t = 3.0\times 10^{-3}$. We run all simulations for $2.0 \times 10^6$ timesteps to ensure that the total kinetic energy of the system has stabilized near a constant value, as verified by monitoring the total energy of the system. Initially, we take all agents to be randomly distributed with zero initial velocities. Having chosen values for all previously discussed parameters, the remaining free parameters are $\phi$, $\alpha$, and $\beta$.

\begin{figure}[!t]
\begin{center}
\subfloat[Mesoturbulent \label{subfig-2:dummy}]{%
\includegraphics[width=.15\textwidth] {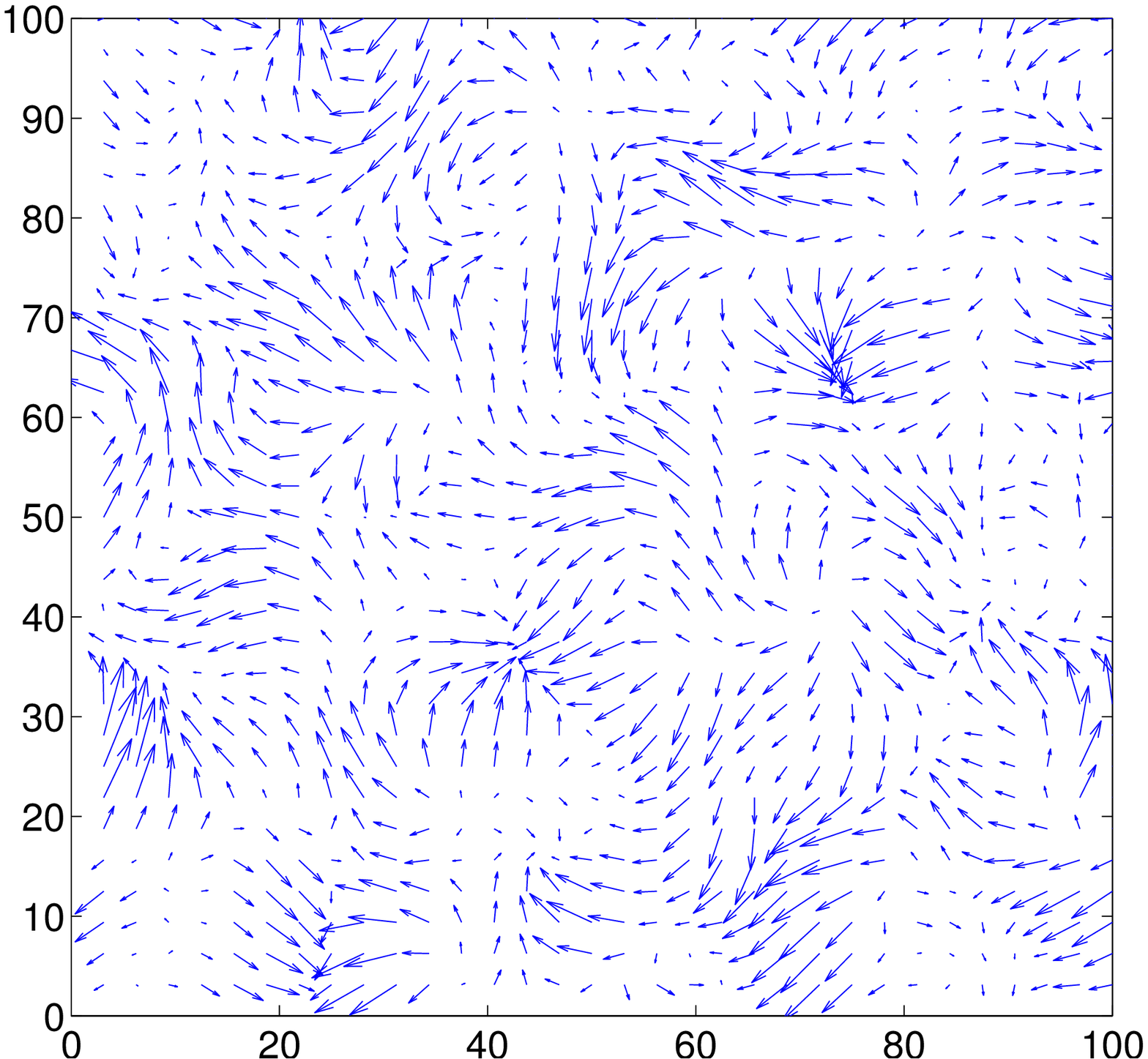}}
\put(-69,61){\fcolorbox{black}{white}{\tiny $\phi = 0.1$, ${\rm Pe} = 0.11$}}
\subfloat[Polar flock \label{subfig-1:dummy}]{%
\includegraphics[width=.15\textwidth] {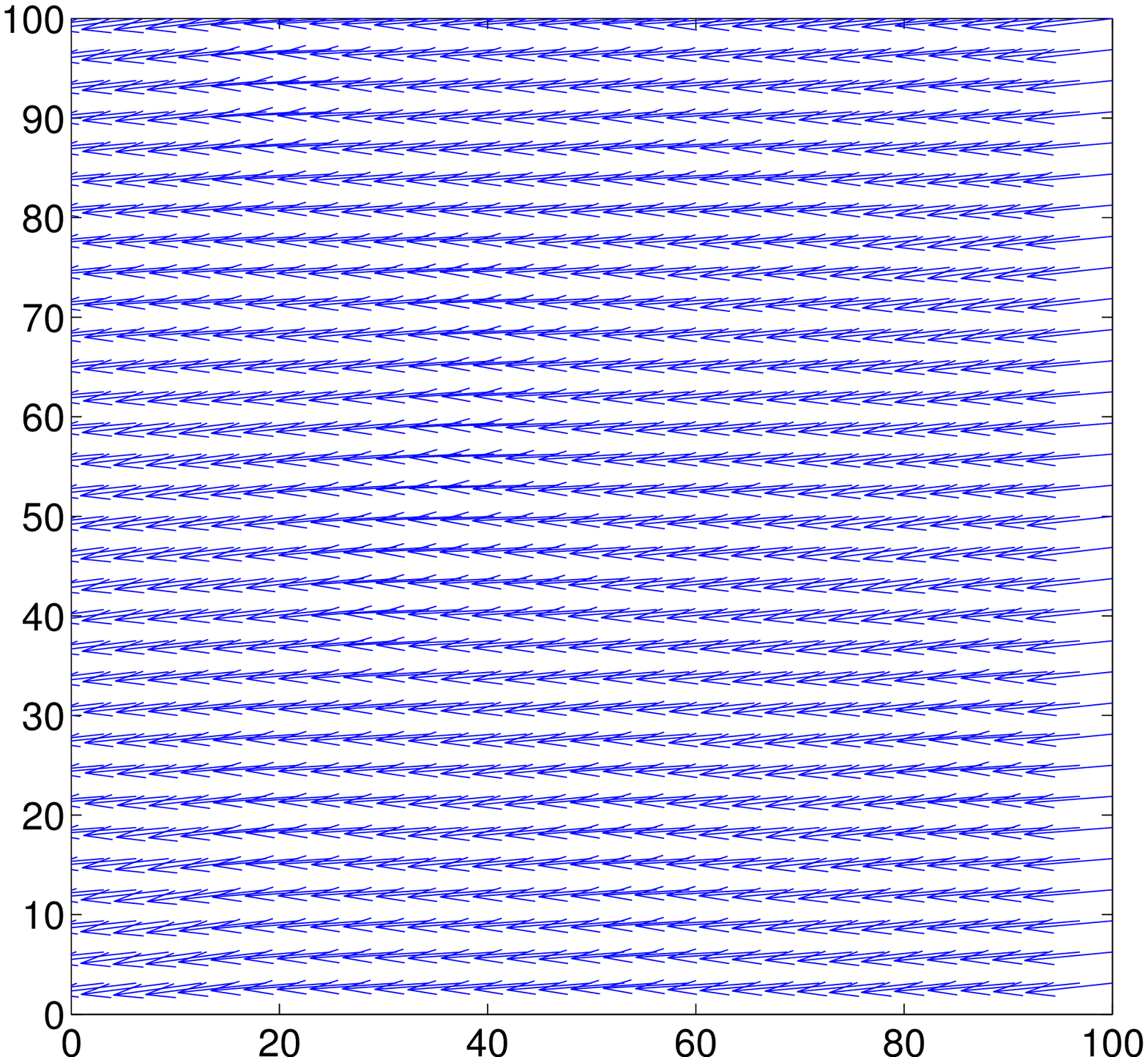}}
\put(-69,61){\fcolorbox{black}{white}{\tiny $\phi = 0.5$, ${\rm Pe} = 1.11$}}
\subfloat[Vortical \label{subfig-1:dummy}]{%
\includegraphics[width=.15\textwidth] {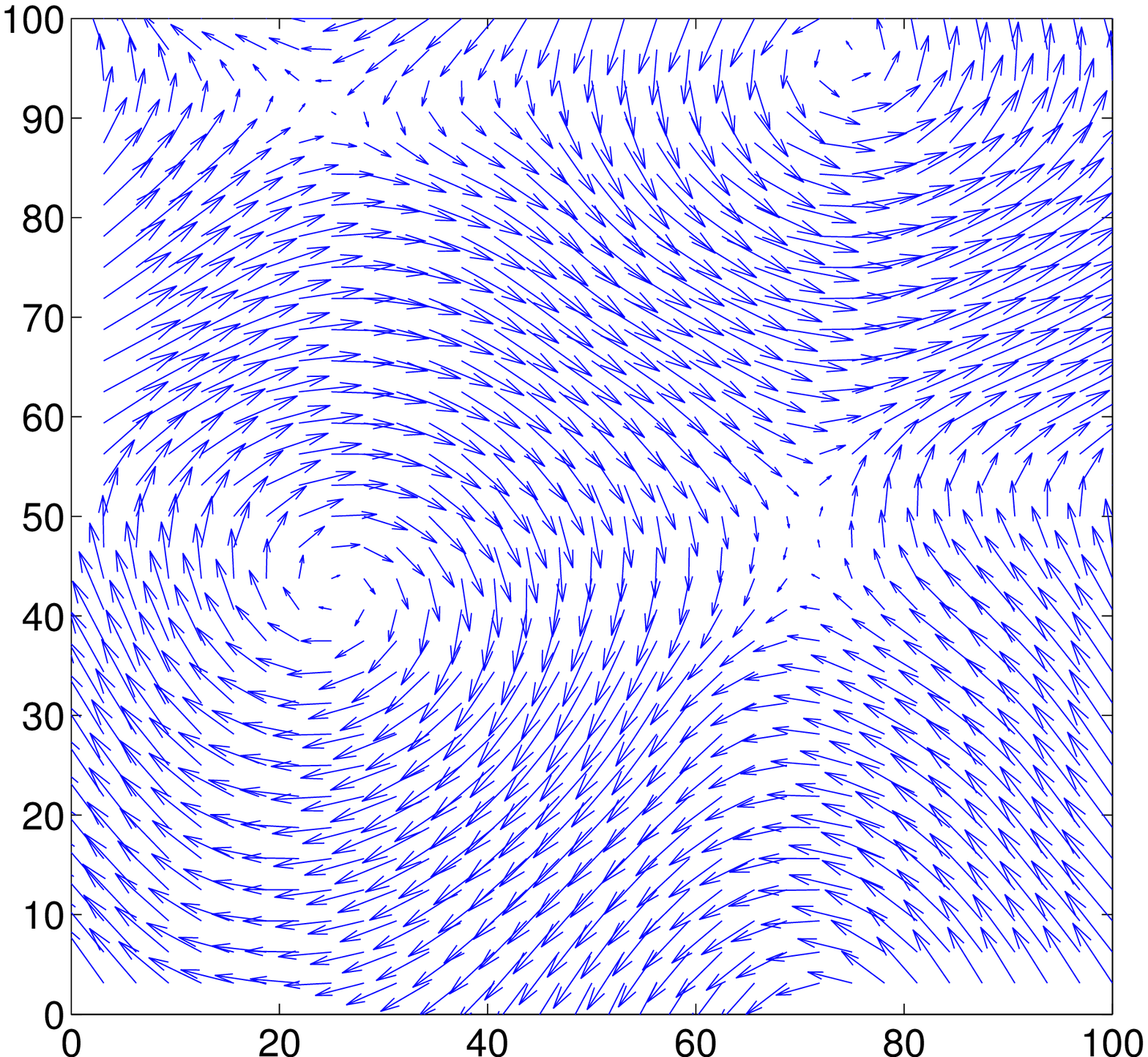}}
\graphicspath{ {./graphics/snapshots/serie14b2p25/T_1p0/alpha_10p0_delta2_4p5_actFrac_0p1//} }
\put(-69,61){\fcolorbox{black}{white}{\tiny $\phi = 0.1$, ${\rm Pe} = 1.11$}}
%
\end{center}
\caption{Coarse-grained velocity field $\bfu_M$ for $\beta = 2.25$ and representative choices of $\phi$ and $\rm{Pe}$. (a) arises for small $\phi$ in combination with small $\rm{Pe}$; (b) arises for intermediate and large $\phi$ in combination with large $\rm{Pe}$; (c) arises for small $\phi$ and $\rm{Pe}>1$. 
}
\label{fig:velField01}
\end{figure}

\begin{table}
\caption{Summary of criteria used to identify different phases. 
}
\label{t:criteria}
 \begin{ruledtabular}
 \begin{tabular}{llll}
Phase 			&$P_v$ 				&$P_\omega$			& snapshot				\\
Mesoturbulent (T) 		&$P_v \rightarrow 1$ 	&$P_\omega \rightarrow 0$		& Figure~\ref{fig:velField01} (a) \\
Polar flock (F)    		&$P_v \rightarrow 0$ 	&$P_\omega \rightarrow 0$		 & Figure~\ref{fig:velField01} (b) \\
Vortical (V) 		&$P_v \rightarrow 1$ 	&$P_\omega > 0$					& Figure~\ref{fig:velField01} (c) \\
 \end{tabular}
 \end{ruledtabular}
  \end{table}

\paragraph{Phase diagram.}
\label{sec:meas}

\begin{figure}[!t]
\hspace{3.6cm}
\includegraphics[width=.22\textwidth] {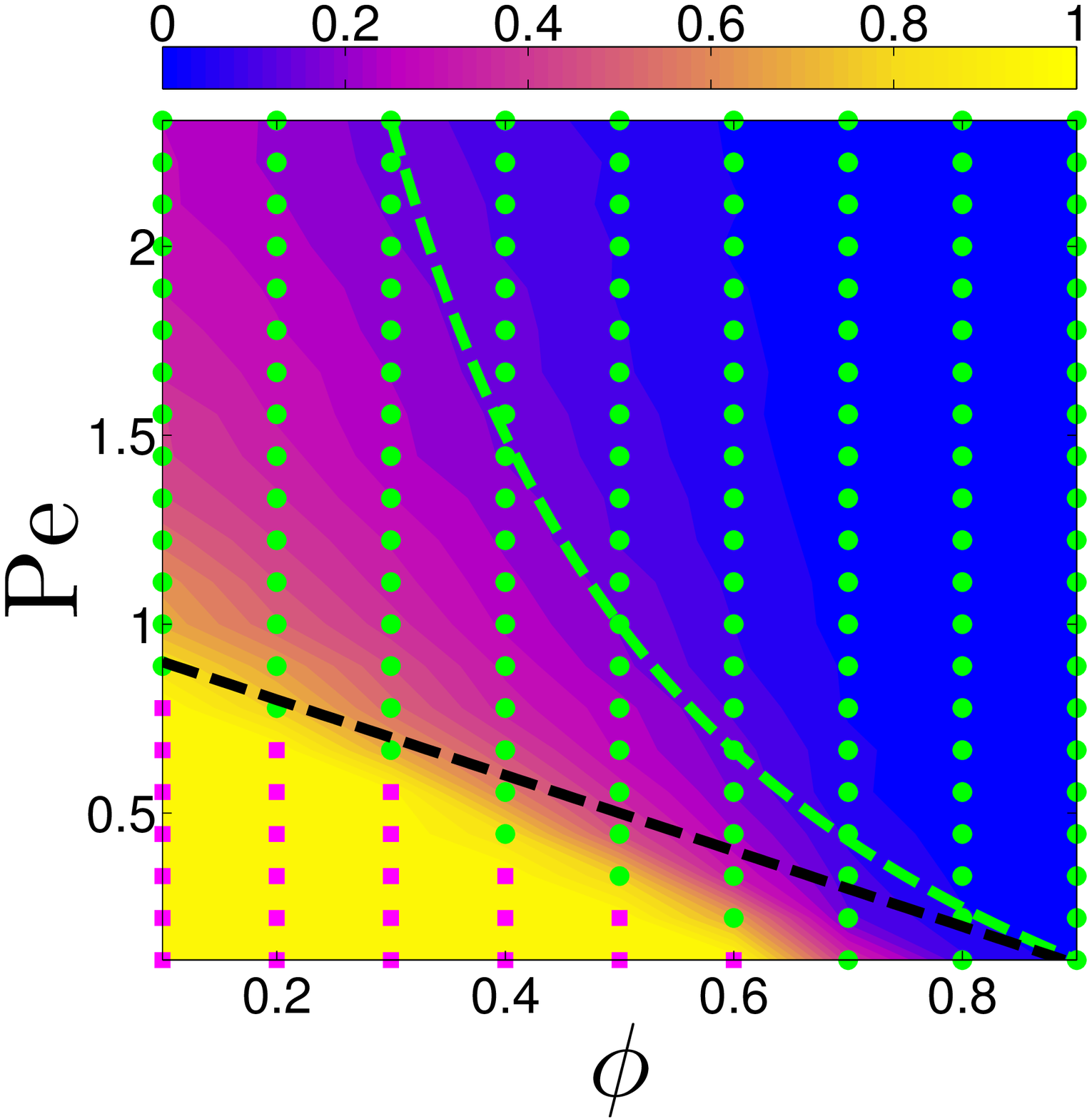}
\put(-110,105){\large$P_v$}
\put(-45,93){\fcolorbox{black}{white}{\tiny$L=25.0r_c$}}
\put(-23,76){\fcolorbox{black}{white}{(b)}}
\put(-230,0){\includegraphics[width=.22\textwidth] {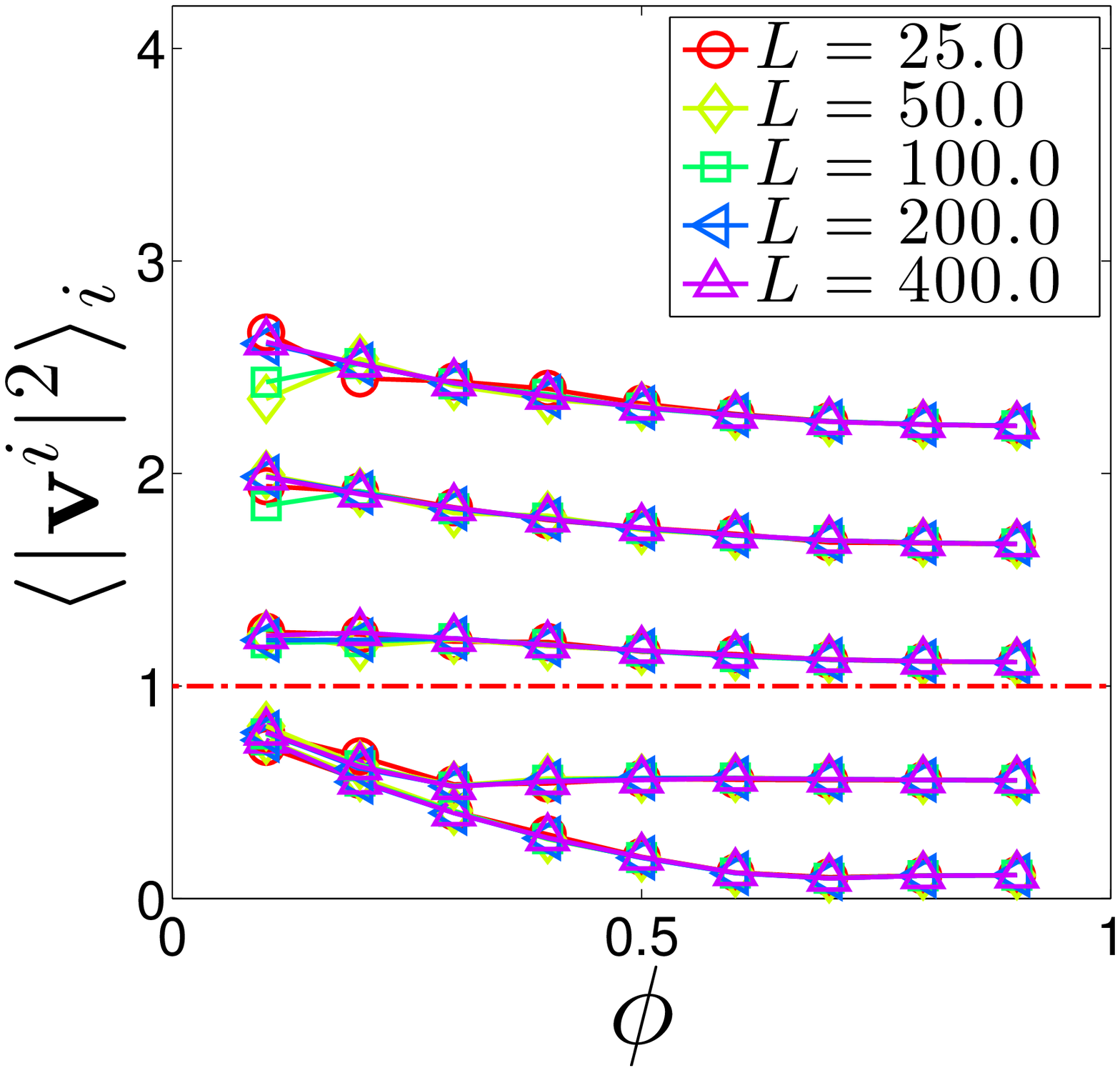} }
\put(-150,68){\tiny${\rm{Pe} = 2.22}$}
\put(-150,55){\tiny${\rm{Pe} = 1.67}$}
\put(-150,43){\tiny${\rm{Pe} = 1.11}$}
\put(-150,32){\tiny${\rm{Pe} = 0.56}$}
\put(-150,22){\tiny${\rm{Pe} = 0.11}$}
\put(-205,90){\fcolorbox{black}{white}{(a)}}

\hspace{3.6cm}
\includegraphics[width=.22\textwidth] {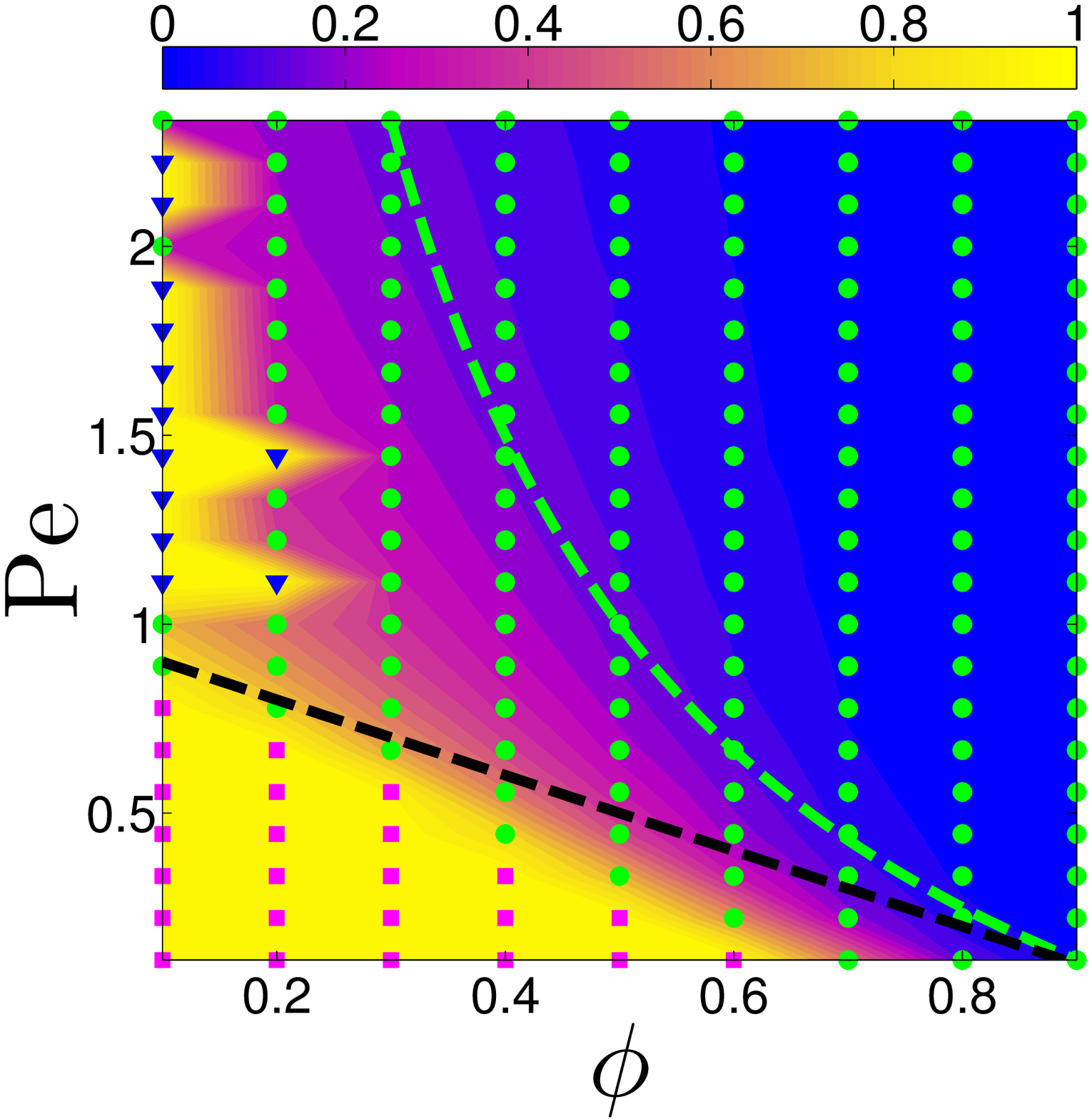}
\put(-110,105){\large$P_v$}
\put(-48,93){\fcolorbox{black}{white}{\tiny$L=100.0r_c$}}
\put(-23,76){\fcolorbox{black}{white}{(d)}}
\put(-230,0){\includegraphics[width=.22\textwidth] {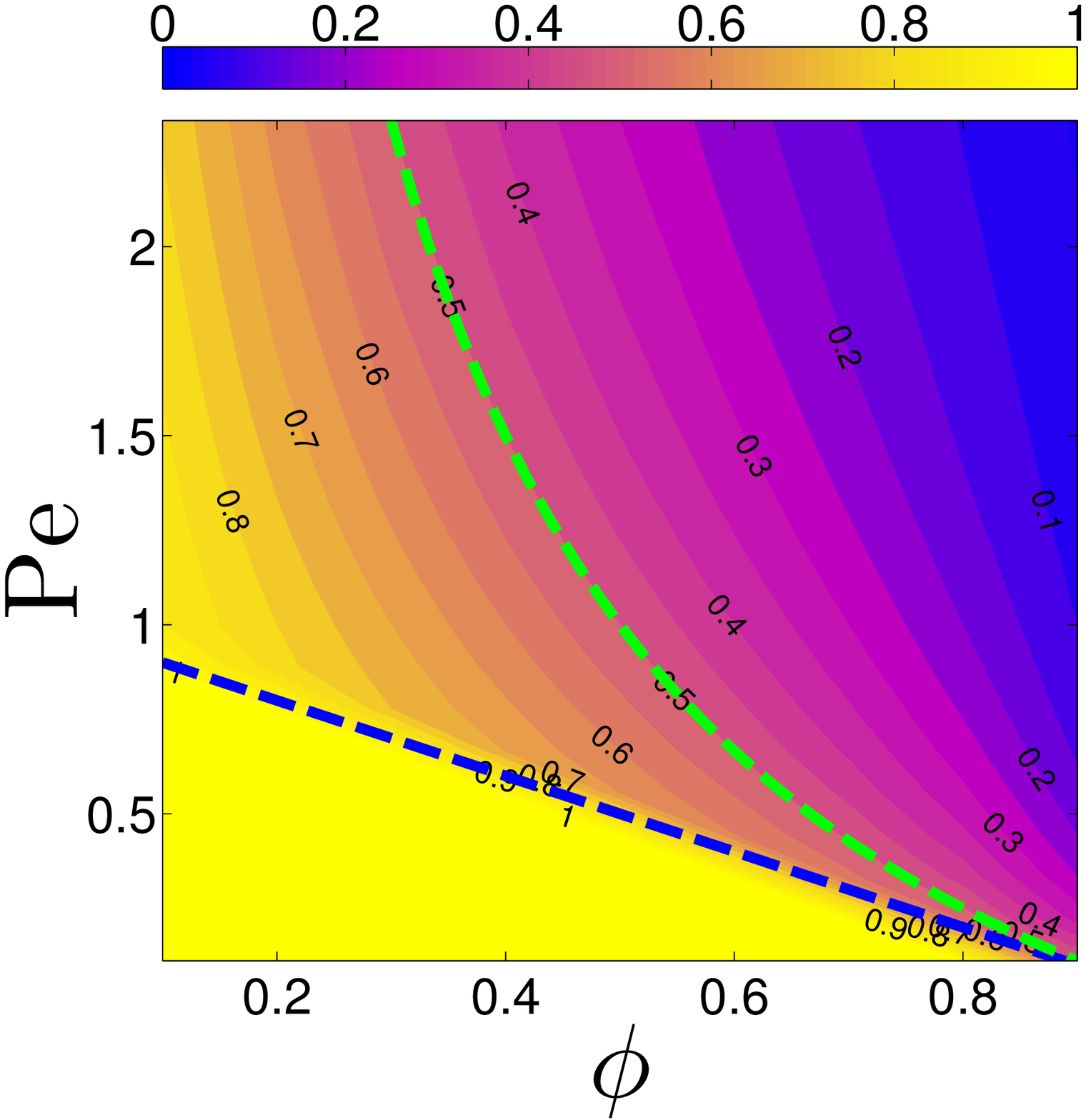} }
\put(-228,105){\large$P_v$}
\put(-141,90){\fcolorbox{black}{white}{(c)}}
\put(-210,20){\color{blue}(T)}
\put(-150, 45){\color{blue}(F)}
\put(-212, 70){\color{blue}(V)}
\put(-196,32){\rotatebox{-16}{$\rm{Pe}=1-\phi$}}
\put(-189,98){\rotatebox{-62}{$\rm{Pe}=(1-\phi)/\phi$}}

\hspace{3.5cm}
\includegraphics[width=.22\textwidth] {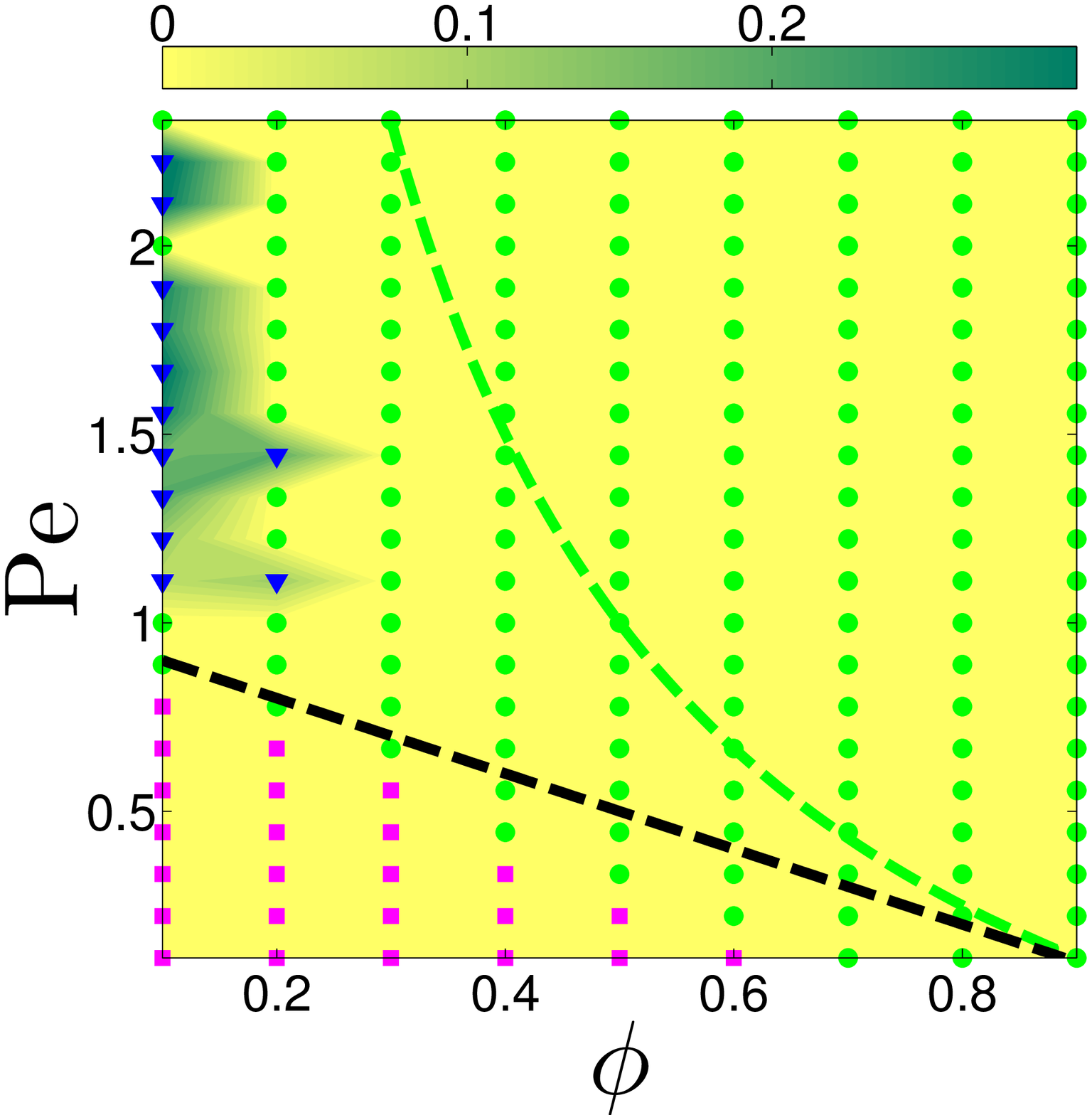}
\put(-110,105){\large$P_\omega$}
\put(-48,93){\fcolorbox{black}{white}{\tiny$L=100.0r_c$}}
\put(-23,76){\fcolorbox{black}{white}{(i)}}
\put(-221,60){\includegraphics[width=.105\textwidth, trim=7cm 7cm 7cm 7cm, clip=true] {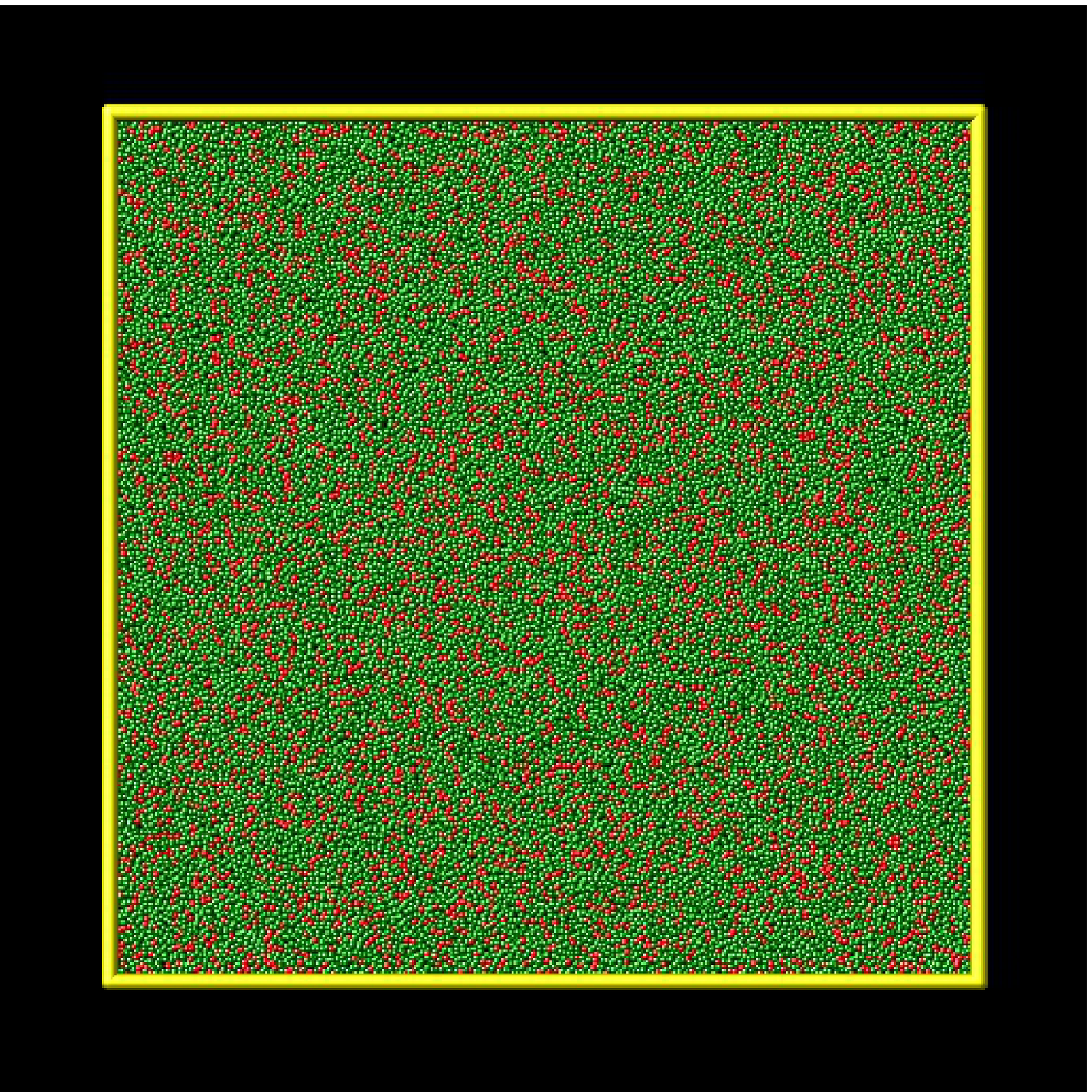}}
\put(-166,60){\includegraphics[width=.105\textwidth, trim=7cm 7cm 7cm 7cm, clip=true] {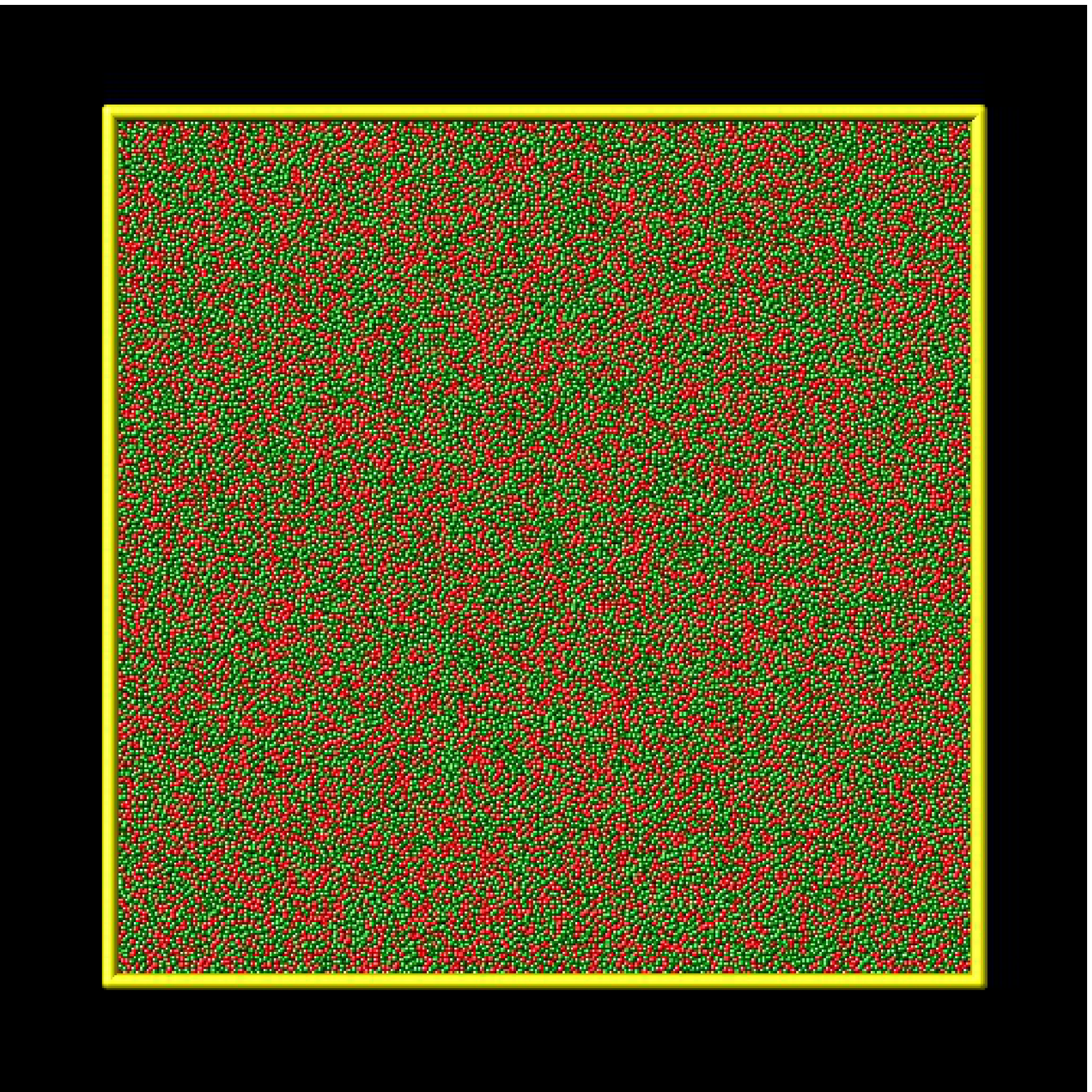}}
\put(-221,5){\includegraphics[width=.105\textwidth, trim=7cm 7cm 7cm 7cm, clip=true] {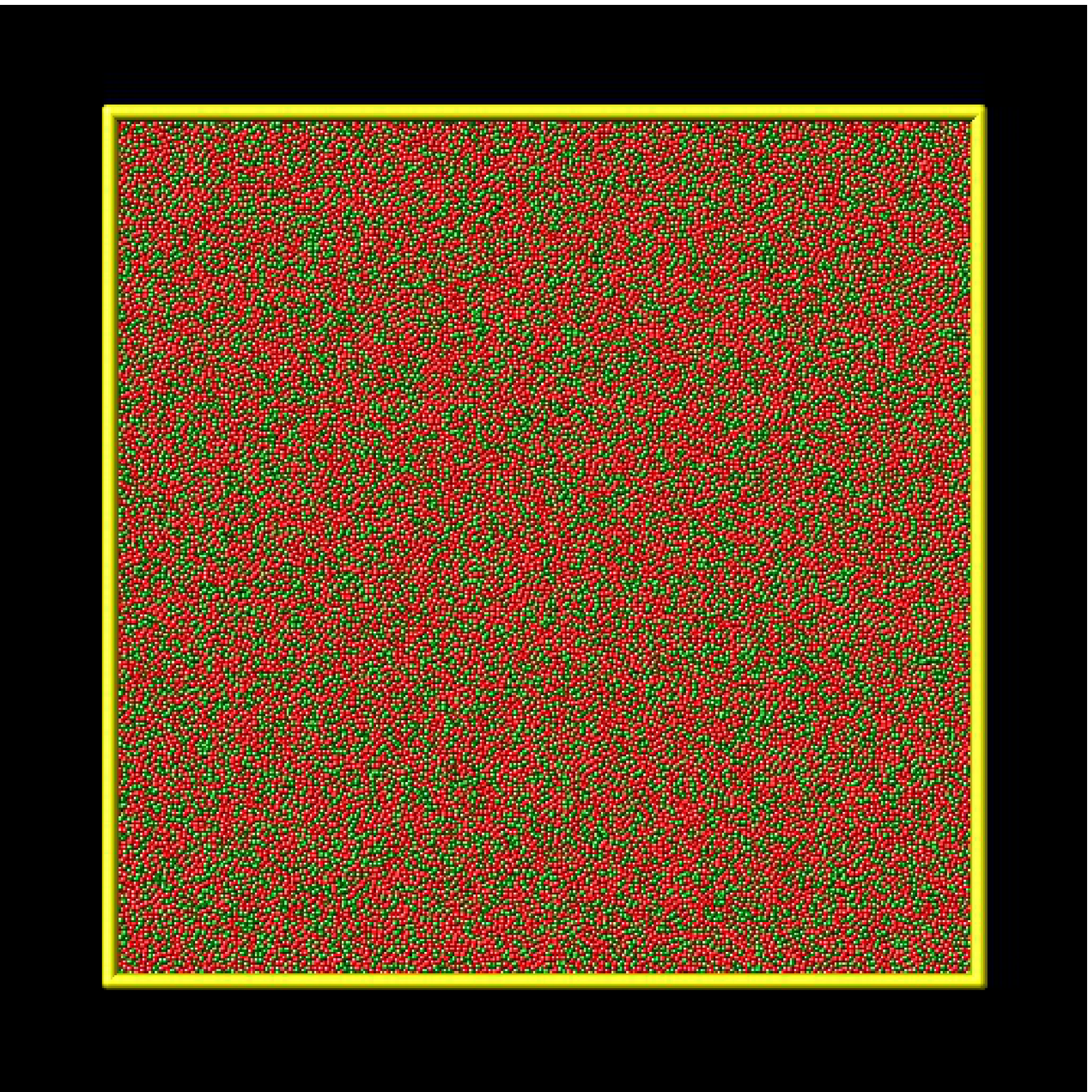}}
\put(-166,5){\includegraphics[width=.105\textwidth, trim=7cm 7cm 7cm 7cm, clip=true] {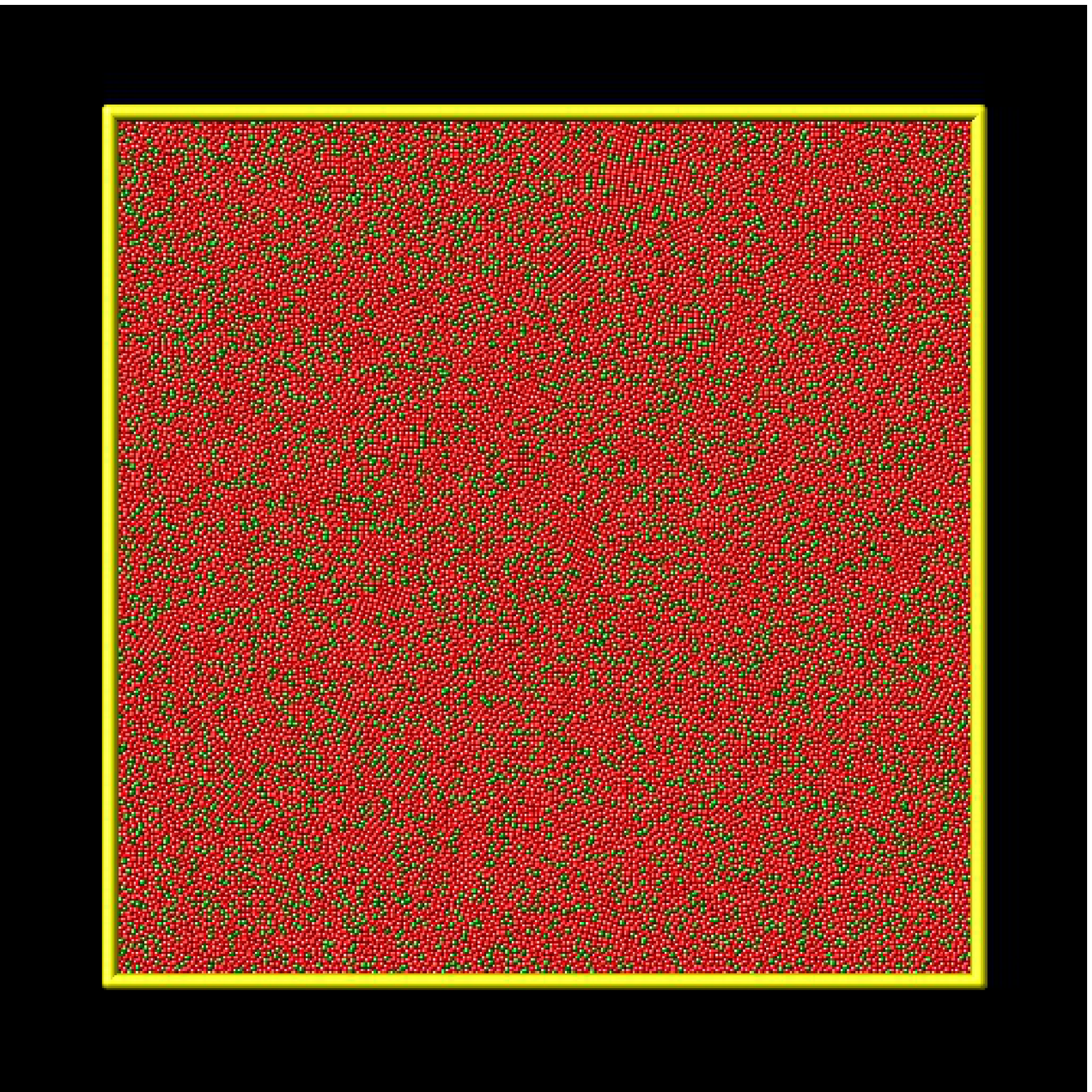}}
\put(-220,102){\fcolorbox{black}{white}{(e) $\phi=0.2$}}
\put(-220,47 ){\fcolorbox{black}{white}{(g) $\phi=0.6$}}
\put(-165,102 ){\fcolorbox{black}{white}{(f) $\phi=0.4$}}
\put(-165,47 ){\fcolorbox{black}{white}{(h) $\phi=0.8$}}
%
\caption{Influence of the domain size, phase diagrams, and agent distribution fields. (a) Total kinetic energy $\langle |\bfv^i|^2 \rangle_i$ for different values of the linear domain size $L$. (b) $P_v$ from simulations ($L=25.0$). (c) $P_v$ from theoretical estimates~\eqref{eq:critTurbND}, \eqref{eq:Pv03}, and~\eqref{eq:Pe_crit04}. (d) $P_v$ from simulations ($L=100.0$). (i) $P_\omega$ from simulations ($L=100.0$). The symbols {\color{magenta}$\blacksquare$}, {\color{blue}$\blacktriangledown$}, and {\color{green}$\bullet$} distinguish, respectively, the mesoturbulent (T), the vortical (V), and the polar flocking (F) phases, as identified by the criteria appearing in Table~\ref{t:criteria}. (e)--(h) Agent distribution fields. }
\label{fig:OrderParam}
\end{figure}

\begin{figure*}[!t]
\begin{center}
\subfloat[Mesoturbulent phase ($\rm{Pe}=0.11$). \label{subfig-2:dummy}]{%
\includegraphics[width=.32\textwidth] {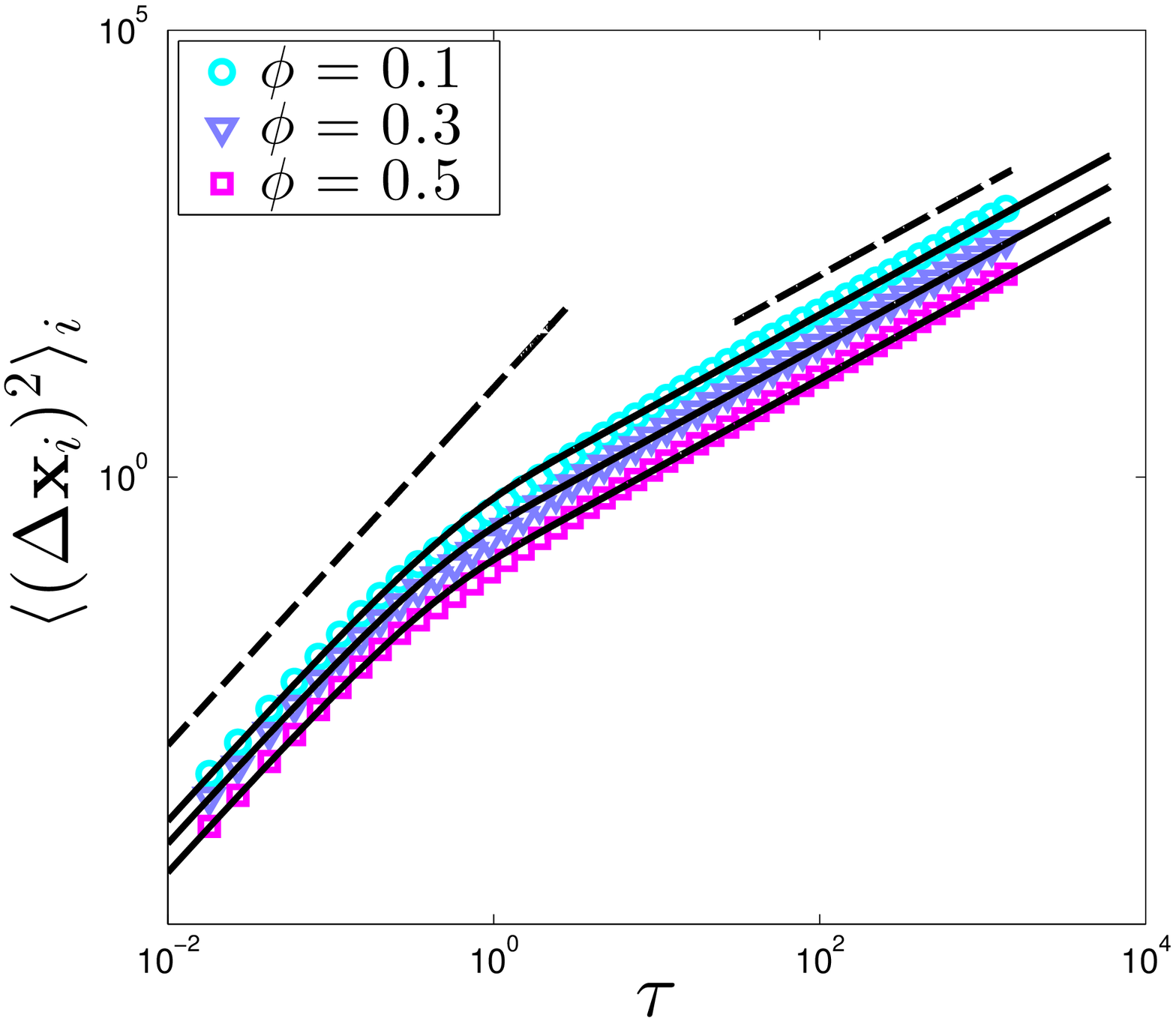}}
\put(-50,115){$\sim  \tau^1$}
\put(-130,70){$\sim  \tau^2$}
\put(-80,15){\includegraphics[width=.14\textwidth] {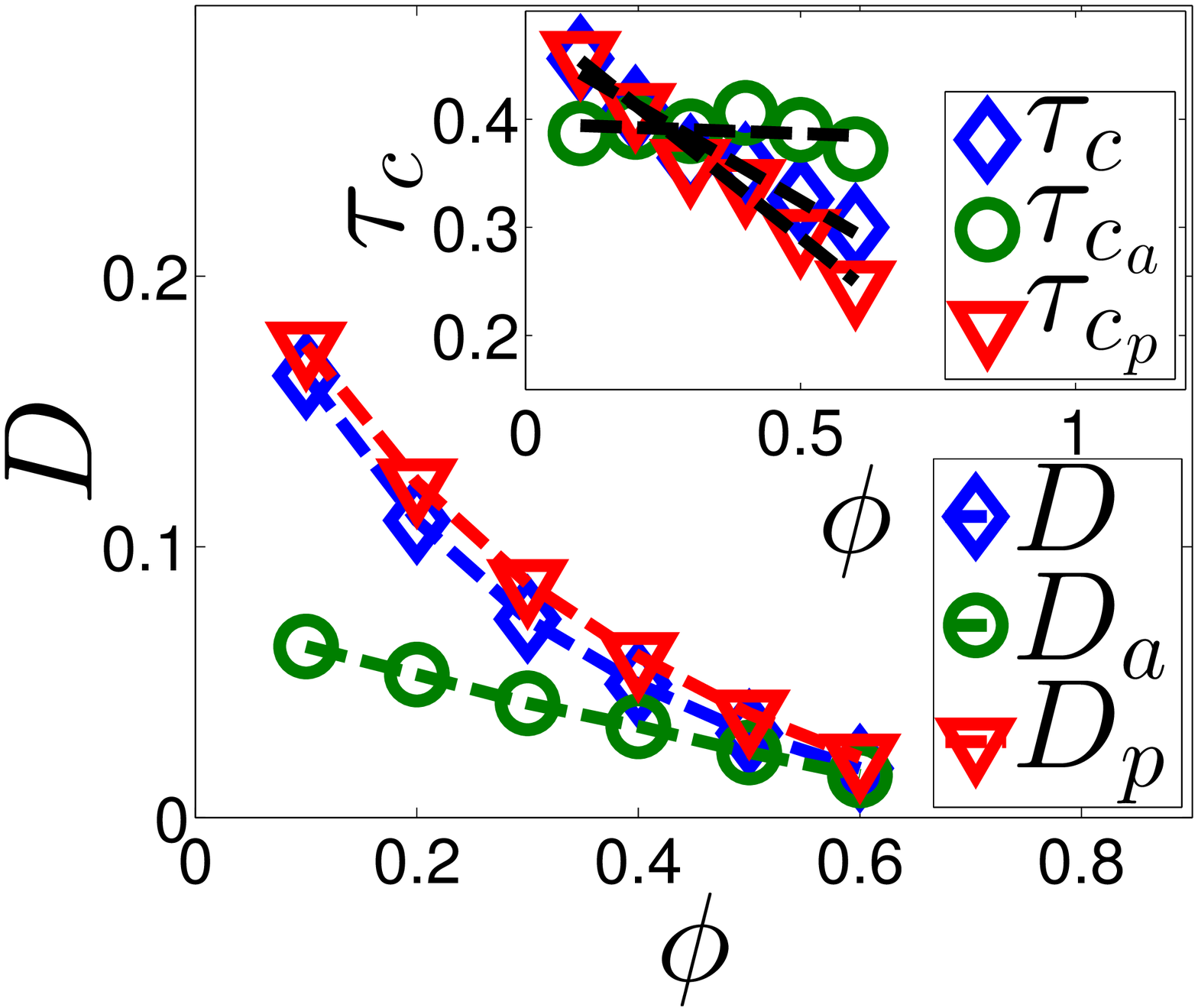}}
\put(-30,78){(a.I)}
\subfloat[Flocking phase ($\rm{Pe}=0.78$). \label{subfig-2:dummy}]{%
\includegraphics[width=.32\textwidth] {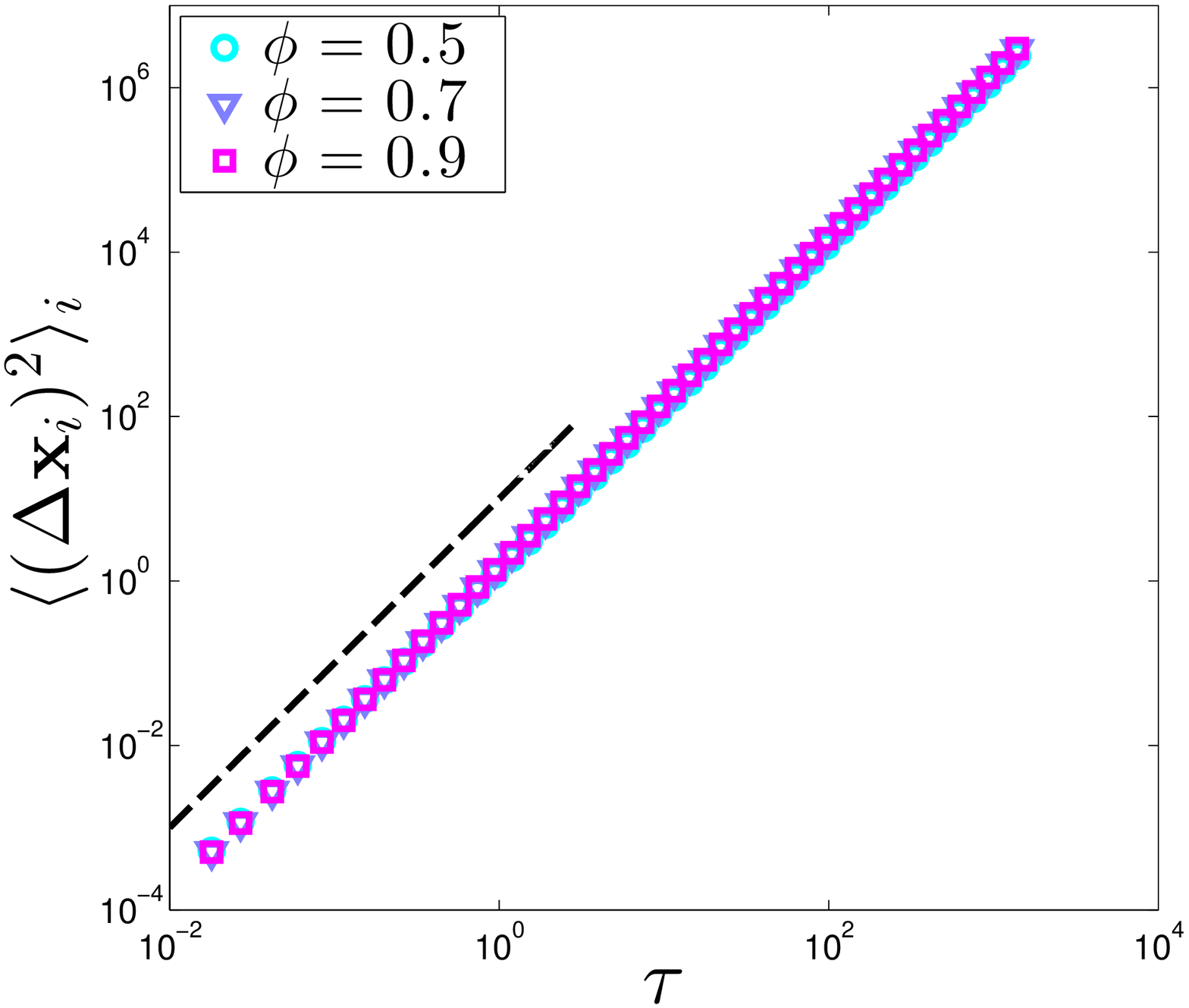}}
\put(-120,70){$\sim  \tau^2$}
\put(-70,15){\includegraphics[width=.125\textwidth] {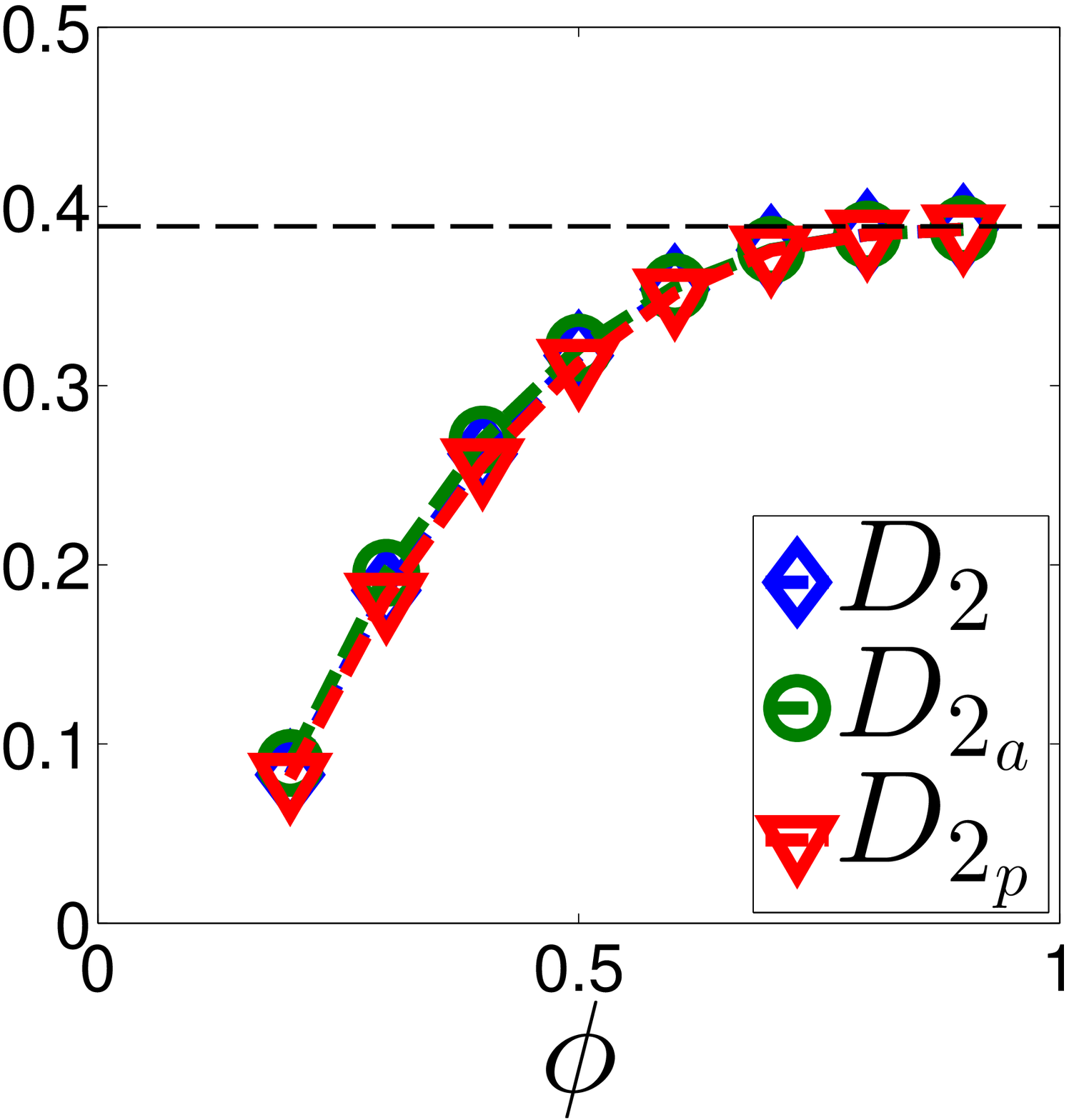}}
\put(-60,58){$\frac{\alpha}{4\beta}$}
\put(-80,50){\rotatebox{90}{$D_2$}}
\put(-30,83){(b.I)}
\subfloat[Vortical phase ($\phi=0.1$). \label{subfig-2:dummy}]{%
\includegraphics[width=.32\textwidth] {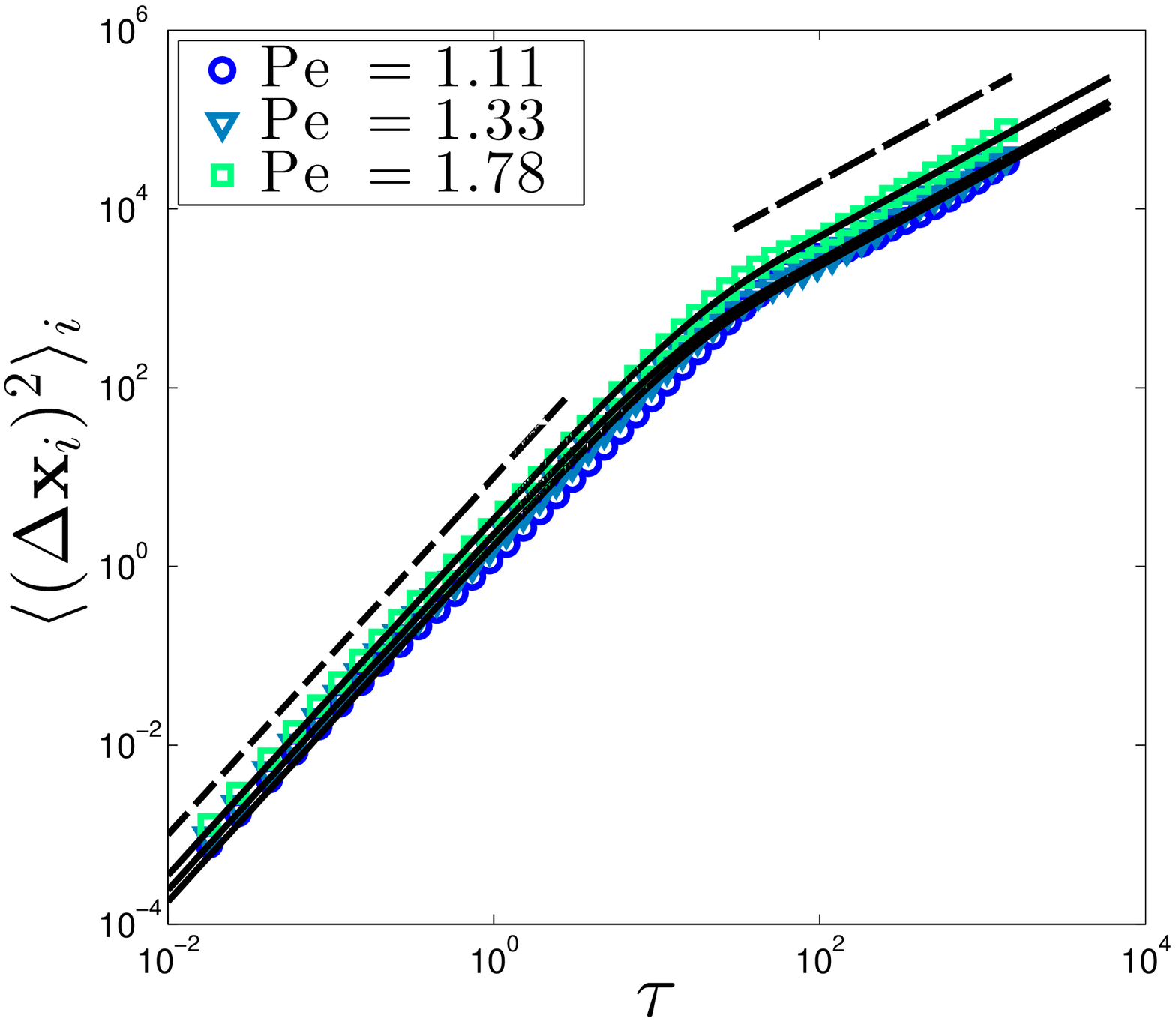}}
\put(-55,125){$\sim  \tau^1$}
\put(-138,50){$\sim  \tau^2$}
\put(-82,15){\includegraphics[width=.15\textwidth] {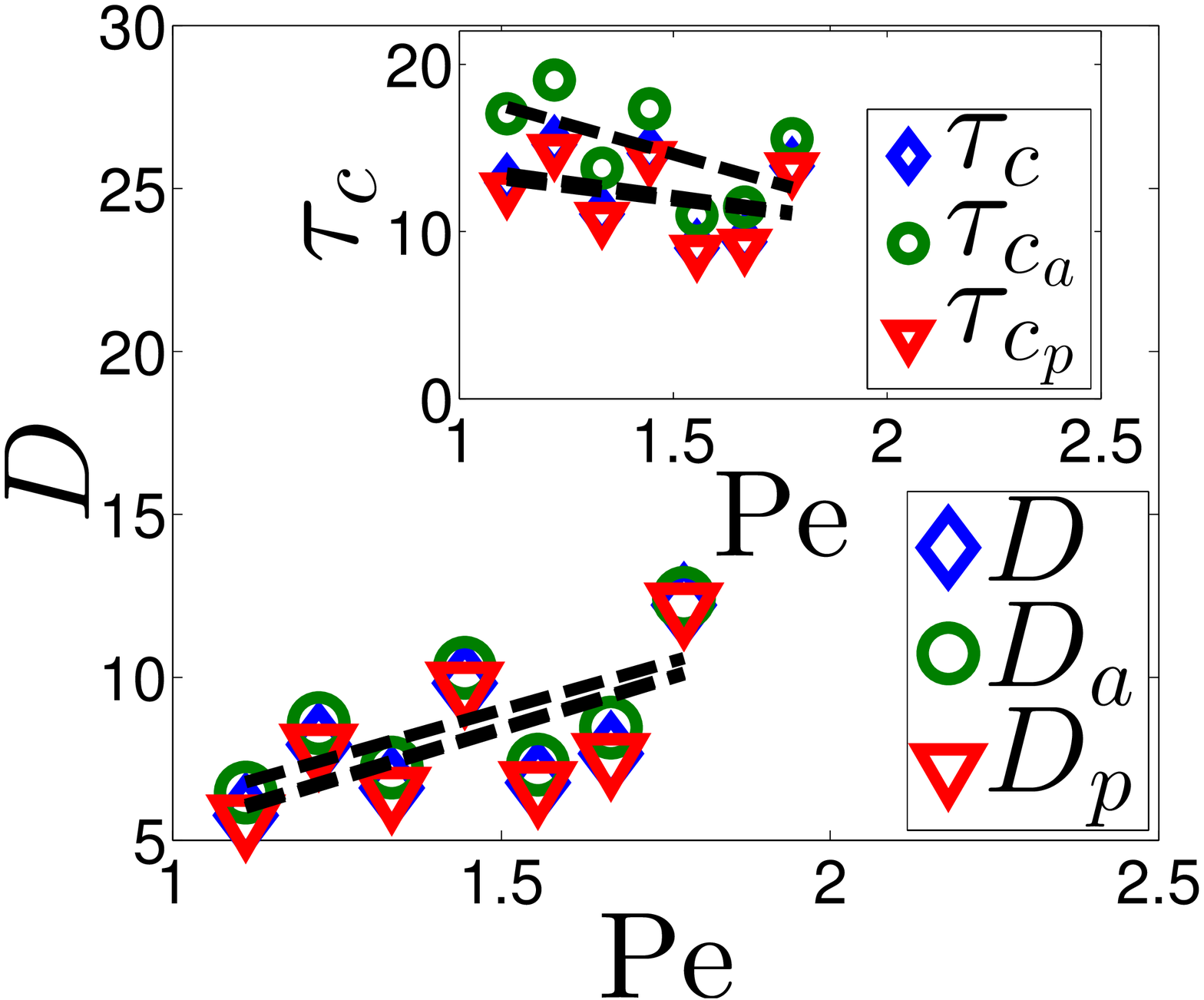}}
\put(-130,76){\includegraphics[width=.075\textwidth] {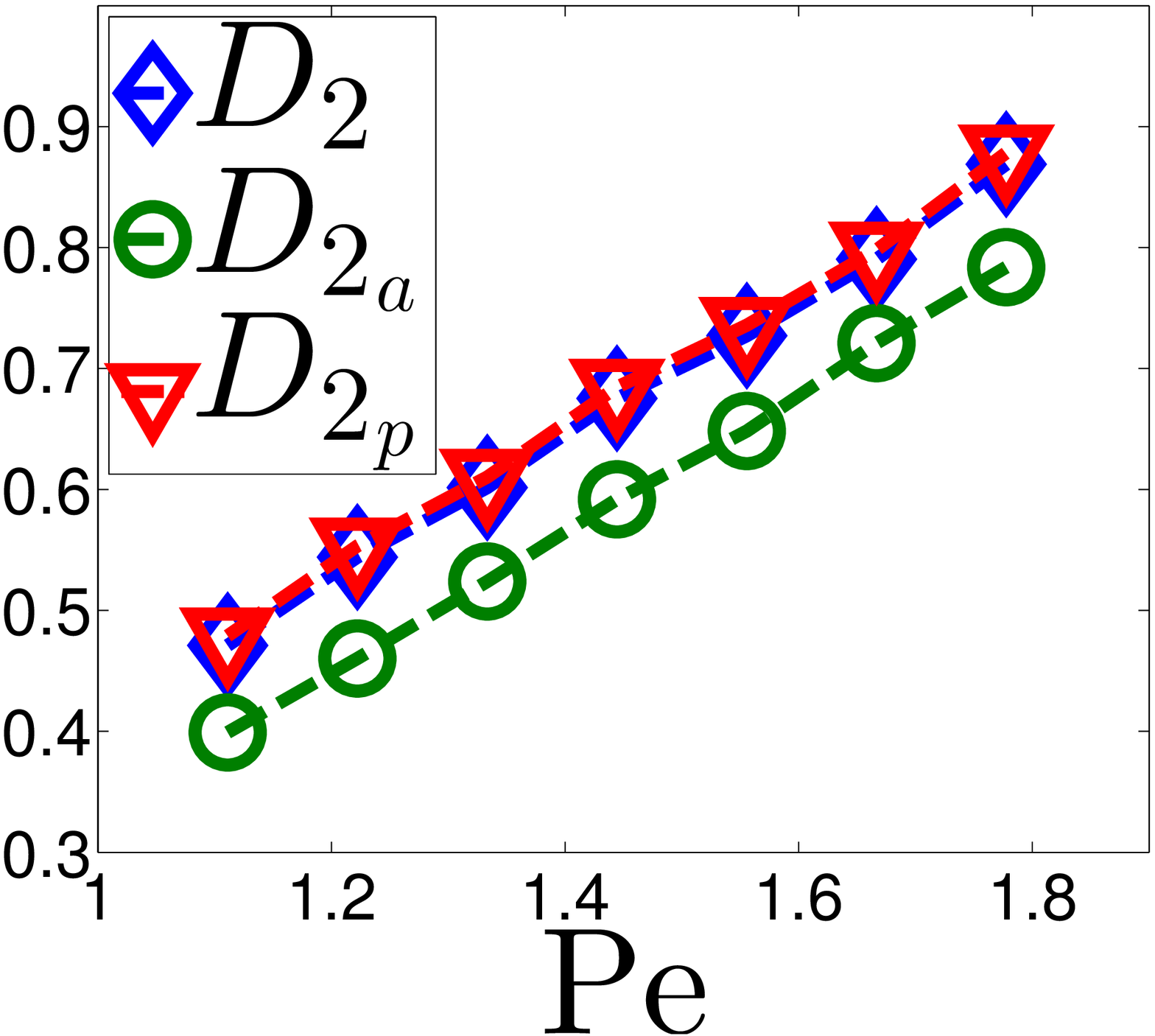}}
\put(-138,90){\rotatebox{90}{$D_2$}}
\put(-30,82){(c.I)}
\put(-92,103){(c.II)}
\end{center}
\caption{MSD for the three different phases. $D_a$ and $t_{c_a}$ ($D_p$ and $t_{c_p}$) denote the parameters associated with the active (passive) fraction of the mixture. (a) MSD for three different fractions of active agents (markers) along with least-squares fit to~\eqref{msddiffWuFit01} indicated through solid lines. (a.I) Diffusion coefficients and crossover time  obtained from a fit of the data to \eqref{msddiffWuFit01}. (b) MSD for three different fractions of active agents. (b.I) Ballistic coefficients obtained from the relation \eqref{msddiff01} with $\xi=2$. (c) MSD along with least-squares fit to~\eqref{msddiffWuFit01} indicated through solid lines. (c.I) Short-time ballistic coefficients obtained from the relation \eqref{msddiff01} with $\xi=2$ and  (c.II) long-time diffusion coefficients as well as crossover time obtained from a fit of the data to \eqref{msddiffWuFit01}. }
\label{fig:MSDturb01}
\end{figure*}

Let 
\begin{equation}\label{eq:Pe}
{\rm{Pe}}=  \frac{ \alpha}{ 2\beta k_B T}
\end{equation}
denote a dimensionless P\'eclet number characterizing the ratio of the self-propulsion energy to the energy of random fluctuations. Depending on the values of $\rm{Pe}$ and $\phi$, the system develops three different phases: a disordered mesoturbulent phase, a polar flocking phase, and a vortical phase characterized by large-scale counterrotating vortices (Fig.~\ref{fig:velField01}). Importantly, no segregation is observed (Fig.~\ref{fig:OrderParam} (e)--(h)). We use order parameters based on the agent velocities $\bfv^i$ and the coarse-grained vorticity $\bfomega = \Curl \bfu_M$ to quantify the influence of $\phi$ and $\rm{Pe}$ on phase emergence. The coarse-grained velocity $\bfu_M(\bfx_M) = \frac{1}{n} \sum_{i=1}^N \bfv^{i}\psi(\bfx_M - \bfx^{i})$ is computed on a uniform grid with equidistant spacing in both coordinate directions, where $\psi$ is a sufficiently rapidly decaying filtering kernel and $n$ is the number of agents for which $\psi \neq 0$. For simplicity, we use a Gaussian filter with nondimensional filter width $\epsilon = 15.0$. The order parameters $P_v$ and $P_\omega$ are defined as
\begin{equation}\label{eq:Pv}
P_v = \frac{\langle |\bfv^i|^2 \rangle_i - \langle \bfv^i \rangle^2_i}{\langle |\bfv^i|^2 \rangle_i}  \quad \text{and} \quad P_\omega = \langle |\bfomega|^2 \rangle - \langle \bfomega \rangle^2,
\end{equation}
where $\langle \cdot \rangle_i$ denotes the spatial average over all agents in the computational domain and $\langle \cdot \rangle$ denotes the spatial average over all points $\bfx_M$. Whereas $P_v$ distinguishes between polar and non-polar states, $P_\omega$ distinguishes between non-polar phases exhibiting large-scale vortical motion and non-polar mesoturbulent phases. The associated limiting cases are summarized in Table~\ref{t:criteria}. We determine the order parameters numerically and exhibit them as 2D contour plots in the $(\phi, \rm{Pe})$--plane (Fig.~\ref{fig:OrderParam}). The phases in the $(\phi, \rm{Pe})$--plane are marked using the identification criteria summarized in Table~\ref{t:criteria}. To test for finite-size effects, we conducted all sets of simulations with different domain sizes ranging from  $L=25.0r_c$ to $L=400.0r_c$.  The vortical phase arises for small $\phi$ and $\rm{Pe}> 1$ in combination with large enough domain sizes (Fig.~\ref{fig:OrderParam} (d) and (i)), whereas for the smallest domain size $r=25.0r_c$ the vortical phase is absent (Fig.~\ref{fig:OrderParam} (b)). The mesoturbulent phase appears to develop independently of the domain size in a triangular region enclosed by the origin, a point near $\phi \rightarrow 0$ and ${\rm Pe}\approx1$, and a point near $\phi\rightarrow 1$ and ${\rm{Pe}}\rightarrow0$. The total mean kinetic energy is found to coincide for all parameter combinations except those corresponding to the vortical phase (Fig.~\ref{fig:OrderParam} (a)), which leads to the conclusion that the mesoturbulent and flocking phases are independent of the domain size. These results indicate that the vortical phase emerges only when the domain size is large enough to accommodate counterrotating vortices. Previous studies showed that an active suspension can be stabilized into a large-scale circulating state through appropriate confinement~\cite{Woodhouse2012,Wioland2013a}. Further, the vortical phase in our simulations appears to resemble ``milling" patterns observed in investigations of ``pure" systems of self-motile particles~\cite{Levine2000, DOrsogna2006, Chuang2007, Chuang2007a, Carrillo2009, Carrillo2010} with the flocking term~\eqref{eq:flocking01}. Most remarkably, in our simulations the large-scale vortical patterns are found without the inclusion of stabilizing attractive forces or confining boundaries. Since the vortical phase occurs for small $\phi$, the passive particles appear to play an important role in stabilizing vortical motion. This effect might be of key importance and deserves future investigation. However, in this letter we focus on the transition criteria between the mesoturbulent phase and the flocking phase.  Unless otherwise noted, the following discussion refers to simulations performed with $L=100.0r_c$.

Heuristically, the mesoturbulent phase is observed for parameter values of $\phi$ and $\rm Pe$ satisfying 
\begin{equation}\label{eq:crit01}
\rm Pe \lesssim 1-\eta\phi, 
\end{equation}
where $\eta$ is an empirical parameter describing the slope of the line  in the $(\phi, \rm{Pe})$--plane below which the mesoturbulent phase prevails. To understand~\eqref{eq:crit01}, consider the average kinetic energies, per particle,  $E_a$ and $E_p$ corresponding to, respectively, active and passive agents. 
These objects are obtained by dividing the phase-specific average kinetic energies $K_a$ and $K_p$ by the total number of particles $N=N_a+N_p$. 
The energy $E_p$ can be estimated by the thermal equilibrium energy $d/2(1-\phi)k_B T$, where  $d=2$ is the physical space dimension. This energy is compared with the total energy $E_p+E_a$ in the polar flocking state. In this state, every particle has a kinetic energy approximately equal to $\frac{\alpha}{2\beta}$. Thus, for a polar flock, $E_p+E_a\approx  \frac{\alpha}{2\beta}$. Heuristically, the meso-turbulent state emerges when the thermal kinetic energy of the passive particles exceeds the total energy of the flocking state, namely when $(1-\phi)k_B T \gtrsim \frac{\alpha}{2\beta}$ or, equivalently, 
\begin{equation}\label{eq:critTurbND}
{\rm Pe}\lesssim1-\phi.
\end{equation} 
Comparison of~\eqref{eq:crit01} and~\eqref{eq:critTurbND} shows that the results from the numerical simulations agree qualitatively with the theoretical predictions and suggests that $\eta=1$. However, the numerical results (Fig.~\ref{fig:OrderParam} (b) and (d)) indicate that $\eta<1$, resulting in transitions to the flocking phase for values of $\phi$ smaller than theoretically predicted. This effect can be explained by noting that the pairwise dissipative interactions between all agents result in an overall damping of the random fluctuations induced by the passive agents. 

Next, we focus on the values of $P_v$ around the transition between the mesoturbulent and the flocking phases. The contour lines in Fig.~\ref{fig:OrderParam} (b) and (d) indicate a discontinuous transition between these phases, resulting in abrupt changes near a threshold value of $P_v \approx 0.5$. To determine the values of $P_v$ for which transitions between the polar and non-polar states may be expected, consider the estimate of the energy in the polar state in terms of the mean velocity $\langle \bfv^i\rangle_i$ and the fluctuating velocity ${\bfv^i}^\prime = \bfv^i - \langle \bfv^i \rangle_i$, such that, without loss of generality, 
\begin{equation}\label{eq:kinE03}
\half \langle |\bfv^i|^2 \rangle_i = \half \langle \bfv^i \rangle_i^2 + \half  \langle |{\bfv^i}^\prime|^2 \rangle_i. 
\end{equation}
Transitions may be expected when the energies associated with the mean velocity and fluctuations are equal:
\begin{equation}\label{eq:kinE04}
 \langle \bfv^i \rangle_i^2 =   \langle |{\bfv^i}^\prime|^2 \rangle_i. 
\end{equation}
By~\eqref{eq:Pv}$_1$, \eqref{eq:kinE03}, and~\eqref{eq:kinE04}, we thus expect a transition between polar and non-polar phases near $P_v= 0.5$, which is consistent with the numerical results. In the flocking regime, $P_v$ exhibits a nonlinear but continuous dependency on $\phi$ and $\rm{Pe}$ (Fig.~\ref{fig:OrderParam} (b) and (d)). To understand how $P_v$ depends on $\phi$ and $\rm{Pe}$, we estimate values of $P_v$ based on the relevant input parameters and compare them to numerical predictions. In the flocking regime, we estimate the terms entering the definition of the order parameter \eqref{eq:Pv}$_1$ as
$\half \langle |\bfv^i|^2 \rangle_i \approx \phi \frac{\alpha}{2\beta}+(1-\phi)k_B T$ and $\half \langle \bfv^i \rangle_i^2 \approx \phi  \frac{\alpha}{2\beta}$. Consequently, we find that 
\begin{equation}\label{eq:Pv03}
P_v \approx \frac{1-\phi}{\phi \rm{Pe}+1-\phi}.
\end{equation}
Importantly,~\eqref{eq:Pv03} provides an estimate for the order parameter $P_v$ in the flocking regime of the $(\rm{Pe},\phi)$--plane. In view of the previous discussion, the transition between the polar flocking phase and the disordered phase occurs at $P_v\approx 0.5$, yielding a criterion similar to~\eqref{eq:critTurbND}:
\begin{equation}\label{eq:Pe_crit04}
\rm{Pe}\approx \frac{1-\phi}{\phi}.
\end{equation}
The good qualitative agreement between the theoretically estimated phase diagram and the numerical predictions (Fig.~\ref{fig:OrderParam} (b)--(d)) confirms the validity of the criteria~\eqref{eq:critTurbND}, \eqref{eq:Pv03}, and~\eqref{eq:Pe_crit04} derived from the energy estimates.

\paragraph{Mean-square displacement (MSD) and diffusion coefficients. }
\label{sec:msdDiff}

The MSD $\langle(\Delta \bfx_i)^2 \rangle_i = \langle(\bfx_i(\tau)-\bfx_{i_0})^2$ can be related to the diffusion through the Langevin equations. For a 2D system (cf., e.g.,~\cite{Metzler2000})
\begin{equation}\label{msddiff01}
\langle(\Delta \bfx_i)^2 \rangle_i = 4D_\xi \tau^\xi,
\end{equation}
where $D_\xi$ is the coefficient associated with the power-law exponent $\xi$. For classical diffusive Brownian motion $\xi=1$ and $D_1$ is the  diffusion coefficient~\cite{Einstein1905}, whereas for ballistic motion $\xi=2$ and $D_2$ is proportional to a characteristic energy per unit mass. In the mesoturbulent phase, the present system exhibits ballistic motion at short times and diffusive motion at long times (Fig.~\ref{fig:MSDturb01} (a)). In experiments with 2D bacterial baths with tracer particles, Wu and Libchaber~\cite{Wu2000, Gregoire2001a, Wu2001a} observed a similar crossover from superdiffusive motion with $\xi>1$ to diffusive motion with $\xi\approx 1$. They found good fits of their experimental data with
\begin{equation}\label{msddiffWuFit01}
\langle(\Delta \bfx_i)^2 \rangle_i = 4D \tau (1- \exp(-\tau/\tau_c)),
\end{equation}
where $D$ is the diffusion coefficient and  $\tau_c$ is the crossover time between two different asymtotic regimes; for $\tau \ll \tau_c$ the motion is ballistic and for $\tau \gg \tau_c$ the motion is diffusive. Least-squares fits of our simulation results (Fig.~\ref{fig:MSDturb01} (a)) show good agreement with~\eqref{msddiffWuFit01}, suggesting that the mesoturbulent phase exhibits statistical properties similar to those of 2D bacterial suspensions. Further, the diffusivity as well as the crossover time decrease with increasing $\phi$, indicating that the random fluctuations of the passive agents are mainly responsible for the diffusivity of the mixtures. At low $\phi$, the diffusivity associated with the passive agents is significantly higher than the diffusivity associated with the active agents (Fig.~\ref{fig:MSDturb01} (a.I)). This difference in diffusivities decreases for increasing $\phi$, signifying that the considered flocking mechanism effectively removes diffusivity, ultimately resulting in transitions to the polar flocking phase. 

In contrast to the mesoturbulent phase, the flocking phase exhibits ballistic behavior on all timescales (Fig.~\ref{fig:MSDturb01} (b)). Consistent with this, the MSD scales with $\tau^2$. In the flocking phase, $D_2$ exhibits an increase with increasing $\phi$, approaching what appears to be an asymptotic value for large $\phi$ (Fig.~\ref{fig:MSDturb01} (b.I)). To understand this limiting behavior, consider the MSD over time, which may be estimated to be proportional to the characteristic flocking energy per unit mass $\langle(\Delta \bfx_i)^2 \rangle_i/{\tau^2} \approx \frac{\alpha}{\beta}$. In view of~\eqref{msddiff01}, $D_2 \approx \frac{\alpha}{4\beta}$ for ballistic motion in the flocking phase as shown by the dashed line in Fig.~\ref{fig:MSDturb01} (b.I). Further, the ballistic coefficients associated with the active and passive agents are almost identical. This confirms that, in the flocking regime, the flocking term dominates and drives the passive agents.

In the vortical phase, the system exhibits ballistic motion at short and intermediate times and diffusive motion at long times (Fig.~\ref{fig:MSDturb01} (c)). Notice that, in contrast to the data appearing in Fig.~\ref{fig:MSDturb01} (a)--(b), the results in Fig.~\ref{fig:MSDturb01} (c) appear for different representative P\'eclet numbers and $\phi=0.1$, since the vortical phase emerges only for low $\phi$. Similarly as in the mesoturbulent phase, the fit to~\eqref{msddiffWuFit01} provides good agreement with the data (Fig.~\ref{fig:MSDturb01} (c)). The crossover times to diffusive motion are an order of magnitude higher than those of the mesoturbulent phase (Fig.~\ref{fig:MSDturb01} (c.I)). This is consistent with the presence of a pair of counterrotating vortices and suggests that this structure is coherent for much longer times than the small-scale swirls that distinguish the mesoturbulent phase. The short-time ballistic coefficient exhibits a linear increase with increasing $\rm{Pe}$ (Fig.~\ref{fig:MSDturb01} (c.II)). Since the ballistic coefficient measures the characteristic energy per unit mass of the system, this demonstrates that the energy in the vortical phase depends linearly on $\rm{Pe}$, even for small $\phi$.

\paragraph{Discussion. }

Our results have key implications for both heterogeneous biological systems and heterogeneous artificial active matter systems. Artificial realizations of active matter systems include, for example, catalytically driven Janus particles~\cite{Paxton2005, Mano2005, Gibbs2009, Jiang2010, Volpe2011}, polymer-based nanomotors~\cite{Tao2009, Kapral2013, Wu2013}, and robotic swarms~\cite{Giomi2013, Li2013}.
First, the results suggest that all dynamical phases and the associated motion patterns may be achieved using a relatively low fraction of self-motile agents. In scenarios where active matter is realized with artificial self-motile agents, this implies that a small number of potentially difficult to manufacture and costly agents should suffice to drive passive agents and generate large-scale flow patterns. Further, mixtures of active and passive particles have great potential for microfluidic tasks such as pumping used in combination with confining geometries including, for example, ratchets \cite{Ghosh2013}.

In a different scenario, the transition between mesoturbulent and the polar flocking phases might be controlled by adjusting the fraction of active agents, while fixing  parameters associated with self-propulsion. Technologically, this means that the emergence of either phase could be controlled by switching identical agents on and off to adjust the fraction of active agents. Such a strategy would make it unnecessary to adjust the parameters related to the magnitude of the self-propulsion. The principle of controlling patterns of motion through using different species of active and passive particles has great potential to increase efficiency in self-powered drug delivery systems~\cite{Patra2013}, water purification~\cite{Soler2013a}, and a multitude of microfluidic processes. For example, the concentration of active agents could be used as a simple switch to control the onset of directed transport or pumping through a dynamical phase transition from the mesoturbulent phase to the flocking phase. This potentially promising avenue remains to be investigated experimentally. 

We thank Pinaki Chakraborty, Gustavo Gioia, Russell Todres, and Steven Aird for helpful feedback and discussions on previous versions of the paper.



%


\begin{thebibliography}{33}%
\makeatletter
\providecommand \@ifxundefined [1]{%
 \@ifx{#1\undefined}
}%
\providecommand \@ifnum [1]{%
 \ifnum #1\expandafter \@firstoftwo
 \else \expandafter \@secondoftwo
 \fi
}%
\providecommand \@ifx [1]{%
 \ifx #1\expandafter \@firstoftwo
 \else \expandafter \@secondoftwo
 \fi
}%
\providecommand \natexlab [1]{#1}%
\providecommand \enquote  [1]{``#1''}%
\providecommand \bibnamefont  [1]{#1}%
\providecommand \bibfnamefont [1]{#1}%
\providecommand \citenamefont [1]{#1}%
\providecommand \href@noop [0]{\@secondoftwo}%
\providecommand \href [0]{\begingroup \@sanitize@url \@href}%
\providecommand \@href[1]{\@@startlink{#1}\@@href}%
\providecommand \@@href[1]{\endgroup#1\@@endlink}%
\providecommand \@sanitize@url [0]{\catcode `\\12\catcode `\$12\catcode
  `\&12\catcode `\#12\catcode `\^12\catcode `\_12\catcode `\%12\relax}%
\providecommand \@@startlink[1]{}%
\providecommand \@@endlink[0]{}%
\providecommand \url  [0]{\begingroup\@sanitize@url \@url }%
\providecommand \@url [1]{\endgroup\@href {#1}{\urlprefix }}%
\providecommand \urlprefix  [0]{URL }%
\providecommand \Eprint [0]{\href }%
\providecommand \doibase [0]{http://dx.doi.org/}%
\providecommand \selectlanguage [0]{\@gobble}%
\providecommand \bibinfo  [0]{\@secondoftwo}%
\providecommand \bibfield  [0]{\@secondoftwo}%
\providecommand \translation [1]{[#1]}%
\providecommand \BibitemOpen [0]{}%
\providecommand \bibitemStop [0]{}%
\providecommand \bibitemNoStop [0]{.\EOS\space}%
\providecommand \EOS [0]{\spacefactor3000\relax}%
\providecommand \BibitemShut  [1]{\csname bibitem#1\endcsname}%
\let\auto@bib@innerbib\@empty
\bibitem [{\citenamefont {Marchetti}\ \emph {et~al.}(2013)\citenamefont
  {Marchetti}, \citenamefont {Joanny}, \citenamefont {Ramaswamy}, \citenamefont
  {Liverpool}, \citenamefont {Prost}, \citenamefont {Rao},\ and\ \citenamefont
  {Simha}}]{Marchetti2013}%
  \BibitemOpen
  \bibfield  {author} {\bibinfo {author} {\bibfnamefont {M.~C.}\ \bibnamefont
  {Marchetti}}, \bibinfo {author} {\bibfnamefont {J.~F.}\ \bibnamefont
  {Joanny}}, \bibinfo {author} {\bibfnamefont {S.}~\bibnamefont {Ramaswamy}},
  \bibinfo {author} {\bibfnamefont {T.~B.}\ \bibnamefont {Liverpool}}, \bibinfo
  {author} {\bibfnamefont {J.}~\bibnamefont {Prost}}, \bibinfo {author}
  {\bibfnamefont {M.}~\bibnamefont {Rao}}, \ and\ \bibinfo {author}
  {\bibfnamefont {R.~A.}\ \bibnamefont {Simha}},\ }\href {\doibase
  10.1103/RevModPhys.85.1143} {\bibfield  {journal} {\bibinfo  {journal}
  {Rev. Mod. Phys.}\ }\textbf {\bibinfo {volume} {85}},\ \bibinfo
  {pages} {1143} (\bibinfo {year} {2013})}\BibitemShut {NoStop}%
\bibitem [{\citenamefont {van Ditmarsch}\ \emph {et~al.}(2013)\citenamefont
  {van Ditmarsch}, \citenamefont {Boyle}, \citenamefont {Sakhtah},
  \citenamefont {Oyler}, \citenamefont {Nadell}, \citenamefont {D\'{e}ziel},
  \citenamefont {Dietrich},\ and\ \citenamefont {Xavier}}]{VanDitmarsch2013}%
  \BibitemOpen
  \bibfield  {author} {\bibinfo {author} {\bibfnamefont {D.}~\bibnamefont {van
  Ditmarsch}}, \bibinfo {author} {\bibfnamefont {K.~E.}\ \bibnamefont {Boyle}},
  \bibinfo {author} {\bibfnamefont {H.}~\bibnamefont {Sakhtah}}, \bibinfo
  {author} {\bibfnamefont {J.~E.}\ \bibnamefont {Oyler}}, \bibinfo {author}
  {\bibfnamefont {C.~D.}\ \bibnamefont {Nadell}}, \bibinfo {author}
  {\bibfnamefont {E.}~\bibnamefont {D\'{e}ziel}}, \bibinfo {author}
  {\bibfnamefont {L.~E.~P.}\ \bibnamefont {Dietrich}}, \ and\ \bibinfo {author}
  {\bibfnamefont {J.~B.}\ \bibnamefont {Xavier}},\ }\href {\doibase
  10.1016/j.celrep.2013.07.026} {\bibfield  {journal} {\bibinfo  {journal}
  {Cell Rep.}\ }\textbf {\bibinfo {volume} {4}},\ \bibinfo {pages} {697}
  (\bibinfo {year} {2013})}\BibitemShut {NoStop}%
\bibitem [{\citenamefont {Monds}\ and\ \citenamefont
  {O'Toole}(2009)}]{Monds2009}%
  \BibitemOpen
  \bibfield  {author} {\bibinfo {author} {\bibfnamefont {R.~D.}\ \bibnamefont
  {Monds}}\ and\ \bibinfo {author} {\bibfnamefont {G.~A.}\ \bibnamefont
  {O'Toole}},\ }\href {\doibase 10.1016/j.tim.2008.11.001} {\bibfield
  {journal} {\bibinfo  {journal} {Trends Microbiol.}\ }\textbf {\bibinfo
  {volume} {17}},\ \bibinfo {pages} {73} (\bibinfo {year} {2009})}\BibitemShut
  {NoStop}%
\bibitem [{\citenamefont {Hibbing}\ \emph {et~al.}(2010)\citenamefont
  {Hibbing}, \citenamefont {Fuqua}, \citenamefont {Parsek},\ and\ \citenamefont
  {Peterson}}]{Hibbing2010}%
  \BibitemOpen
  \bibfield  {author} {\bibinfo {author} {\bibfnamefont {M.~E.}\ \bibnamefont
  {Hibbing}}, \bibinfo {author} {\bibfnamefont {C.}~\bibnamefont {Fuqua}},
  \bibinfo {author} {\bibfnamefont {M.~R.}\ \bibnamefont {Parsek}}, \ and\
  \bibinfo {author} {\bibfnamefont {S.~B.}\ \bibnamefont {Peterson}},\ }\href
  {\doibase 10.1038/nrmicro2259} {\bibfield  {journal} {\bibinfo  {journal}
  {Nature Rev. Microbiol.}\ }\textbf {\bibinfo {volume} {8}},\ \bibinfo
  {pages} {15} (\bibinfo {year} {2010})}\BibitemShut {NoStop}%
\bibitem [{\citenamefont {Nadell}\ \emph {et~al.}(2010)\citenamefont {Nadell},
  \citenamefont {Foster},\ and\ \citenamefont {Xavier}}]{Nadell2010}%
  \BibitemOpen
  \bibfield  {author} {\bibinfo {author} {\bibfnamefont {C.~D.}\ \bibnamefont
  {Nadell}}, \bibinfo {author} {\bibfnamefont {K.~R.}\ \bibnamefont {Foster}},
  \ and\ \bibinfo {author} {\bibfnamefont {J.~B.}\ \bibnamefont {Xavier}},\
  }\href {\doibase 10.1371/journal.pcbi.1000716} {\bibfield  {journal}
  {\bibinfo  {journal} {PLoS Comput. Biol.}\ }\textbf {\bibinfo
  {volume} {6}},\ \bibinfo {pages} {e1000716} (\bibinfo {year}
  {2010})}\BibitemShut {NoStop}%
\bibitem [{\citenamefont {Xavier}(2011)}]{Xavier2011}%
  \BibitemOpen
  \bibfield  {author} {\bibinfo {author} {\bibfnamefont {J.~B.}\ \bibnamefont
  {Xavier}},\ }\href {\doibase 10.1038/msb.2011.16} {\bibfield  {journal}
  {\bibinfo  {journal} {Mol. Syst. Biol.}\ }\textbf {\bibinfo {volume}
  {7}},\ \bibinfo {pages} {483} (\bibinfo {year} {2011})}\BibitemShut {NoStop}%
\bibitem [{\citenamefont {McCandlish}\ \emph {et~al.}(2012)\citenamefont
  {McCandlish}, \citenamefont {Baskaran},\ and\ \citenamefont
  {Hagan}}]{McCandlish2012}%
  \BibitemOpen
  \bibfield  {author} {\bibinfo {author} {\bibfnamefont {S.~R.}\ \bibnamefont
  {McCandlish}}, \bibinfo {author} {\bibfnamefont {A.}~\bibnamefont
  {Baskaran}}, \ and\ \bibinfo {author} {\bibfnamefont {M.~F.}\ \bibnamefont
  {Hagan}},\ }\href {\doibase 10.1039/c2sm06960a} {\bibfield  {journal}
  {\bibinfo  {journal} {Soft Matter}\ }\textbf {\bibinfo {volume} {8}},\
  \bibinfo {pages} {2527} (\bibinfo {year} {2012})}\BibitemShut {NoStop}%
\bibitem [{\citenamefont {Fu}\ \emph {et~al.}(2012)\citenamefont {Fu},
  \citenamefont {Tang}, \citenamefont {Liu}, \citenamefont {Huang},
  \citenamefont {Hwa},\ and\ \citenamefont {Lenz}}]{Fu2012}%
  \BibitemOpen
  \bibfield  {author} {\bibinfo {author} {\bibfnamefont {X.}~\bibnamefont
  {Fu}}, \bibinfo {author} {\bibfnamefont {L.-H.}\ \bibnamefont {Tang}},
  \bibinfo {author} {\bibfnamefont {C.}~\bibnamefont {Liu}}, \bibinfo {author}
  {\bibfnamefont {J.-D.}\ \bibnamefont {Huang}}, \bibinfo {author}
  {\bibfnamefont {T.}~\bibnamefont {Hwa}}, \ and\ \bibinfo {author}
  {\bibfnamefont {P.}~\bibnamefont {Lenz}},\ }\href {\doibase
  10.1103/PhysRevLett.108.198102} {\bibfield  {journal} {\bibinfo  {journal}
  {Phys. Rev. Lett.}\ }\textbf {\bibinfo {volume} {108}},\ \bibinfo
  {pages} {198102} (\bibinfo {year} {2012})}\BibitemShut {NoStop}%
\bibitem [{\citenamefont {Liu}\ \emph {et~al.}(2011)\citenamefont {Liu},
  \citenamefont {Fu}, \citenamefont {Liu}, \citenamefont {Ren}, \citenamefont
  {Chau}, \citenamefont {Li}, \citenamefont {Xiang}, \citenamefont {Zeng},
  \citenamefont {Chen}, \citenamefont {Tang}, \citenamefont {Lenz},
  \citenamefont {Cui}, \citenamefont {Huang}, \citenamefont {Hwa},\ and\
  \citenamefont {Huang}}]{Liu2011}%
  \BibitemOpen
  \bibfield  {author} {\bibinfo {author} {\bibfnamefont {C.}~\bibnamefont
  {Liu}}, \bibinfo {author} {\bibfnamefont {X.}~\bibnamefont {Fu}}, \bibinfo
  {author} {\bibfnamefont {L.}~\bibnamefont {Liu}}, \bibinfo {author}
  {\bibfnamefont {X.}~\bibnamefont {Ren}}, \bibinfo {author} {\bibfnamefont
  {C.~K.~L.}\ \bibnamefont {Chau}}, \bibinfo {author} {\bibfnamefont
  {S.}~\bibnamefont {Li}}, \bibinfo {author} {\bibfnamefont {L.}~\bibnamefont
  {Xiang}}, \bibinfo {author} {\bibfnamefont {H.}~\bibnamefont {Zeng}},
  \bibinfo {author} {\bibfnamefont {G.}~\bibnamefont {Chen}}, \bibinfo {author}
  {\bibfnamefont {L.-H.}\ \bibnamefont {Tang}}, \bibinfo {author}
  {\bibfnamefont {P.}~\bibnamefont {Lenz}}, \bibinfo {author} {\bibfnamefont
  {X.}~\bibnamefont {Cui}}, \bibinfo {author} {\bibfnamefont {W.}~\bibnamefont
  {Huang}}, \bibinfo {author} {\bibfnamefont {T.}~\bibnamefont {Hwa}}, \ and\
  \bibinfo {author} {\bibfnamefont {J.-D.}\ \bibnamefont {Huang}},\ }\href
  {\doibase 10.1126/science.1209042} {\bibfield  {journal} {\bibinfo  {journal}
  {Science}\ }\textbf {\bibinfo {volume} {334}},\ \bibinfo {pages} {238}
  (\bibinfo {year} {2011})}\BibitemShut {NoStop}%
\bibitem [{\citenamefont {Stricker}\ \emph {et~al.}(2008)\citenamefont
  {Stricker}, \citenamefont {Cookson}, \citenamefont {Bennett}, \citenamefont
  {Mather}, \citenamefont {Tsimring},\ and\ \citenamefont
  {Hasty}}]{Stricker2008}%
  \BibitemOpen
  \bibfield  {author} {\bibinfo {author} {\bibfnamefont {J.}~\bibnamefont
  {Stricker}}, \bibinfo {author} {\bibfnamefont {S.}~\bibnamefont {Cookson}},
  \bibinfo {author} {\bibfnamefont {M.~R.}\ \bibnamefont {Bennett}}, \bibinfo
  {author} {\bibfnamefont {W.~H.}\ \bibnamefont {Mather}}, \bibinfo {author}
  {\bibfnamefont {L.~S.}\ \bibnamefont {Tsimring}}, \ and\ \bibinfo {author}
  {\bibfnamefont {J.}~\bibnamefont {Hasty}},\ }\href {\doibase
  10.1038/nature07389} {\bibfield  {journal} {\bibinfo  {journal} {Nature}\
  }\textbf {\bibinfo {volume} {456}},\ \bibinfo {pages} {516} (\bibinfo {year}
  {2008})}\BibitemShut {NoStop}%
\bibitem [{\citenamefont {McDaniel}\ and\ \citenamefont
  {Weiss}(2005)}]{McDaniel2005}%
  \BibitemOpen
  \bibfield  {author} {\bibinfo {author} {\bibfnamefont {R.}~\bibnamefont
  {McDaniel}}\ and\ \bibinfo {author} {\bibfnamefont {R.}~\bibnamefont
  {Weiss}},\ }\href {\doibase 10.1016/j.copbio.2005.07.002} {\bibfield
  {journal} {\bibinfo  {journal} {Curr. Opin. Biotechnol.}\ }\textbf
  {\bibinfo {volume} {16}},\ \bibinfo {pages} {476} (\bibinfo {year}
  {2005})}\BibitemShut {NoStop}%
\bibitem [{\citenamefont {You}\ \emph {et~al.}(2004)\citenamefont {You},
  \citenamefont {{Sidney Cox III}}, \citenamefont {Weiss},\ and\ \citenamefont
  {Arnold}}]{You2004}%
  \BibitemOpen
  \bibfield  {author} {\bibinfo {author} {\bibfnamefont {L.}~\bibnamefont
  {You}}, \bibinfo {author} {\bibfnamefont {R.}~\bibnamefont {{Sidney Cox
  III}}}, \bibinfo {author} {\bibfnamefont {R.}~\bibnamefont {Weiss}}, \ and\
  \bibinfo {author} {\bibfnamefont {F.~H.}\ \bibnamefont {Arnold}},\ }\href
  {\doibase 10.1038/nature02468.1.} {\bibfield  {journal} {\bibinfo  {journal}
  {Nature}\ }\textbf {\bibinfo {volume} {428}},\ \bibinfo {pages} {868}
  (\bibinfo {year} {2004})}\BibitemShut {NoStop}%
\bibitem [{\citenamefont {Kobayashi}\ \emph {et~al.}(2004)\citenamefont
  {Kobayashi}, \citenamefont {Kaern}, \citenamefont {Araki}, \citenamefont
  {Chung}, \citenamefont {Gardner}, \citenamefont {Cantor},\ and\ \citenamefont
  {Collins}}]{Kobayashi2004}%
  \BibitemOpen
  \bibfield  {author} {\bibinfo {author} {\bibfnamefont {H.}~\bibnamefont
  {Kobayashi}}, \bibinfo {author} {\bibfnamefont {M.}~\bibnamefont {Kaern}},
  \bibinfo {author} {\bibfnamefont {M.}~\bibnamefont {Araki}}, \bibinfo
  {author} {\bibfnamefont {K.}~\bibnamefont {Chung}}, \bibinfo {author}
  {\bibfnamefont {T.~S.}\ \bibnamefont {Gardner}}, \bibinfo {author}
  {\bibfnamefont {C.~R.}\ \bibnamefont {Cantor}}, \ and\ \bibinfo {author}
  {\bibfnamefont {J.~J.}\ \bibnamefont {Collins}},\ }\href {\doibase
  10.1073/pnas.0402940101} {\bibfield  {journal} {\bibinfo  {journal}
  {Proc. Natl. Acad. Sci. U.S.A.}\ }\textbf {\bibinfo {volume} {101}},\ \bibinfo {pages} {8414}
  (\bibinfo {year} {2004})}\BibitemShut {NoStop}%
\bibitem [{\citenamefont {Paxton}\ \emph {et~al.}(2005)\citenamefont {Paxton},
  \citenamefont {Sen},\ and\ \citenamefont {Mallouk}}]{Paxton2005}%
  \BibitemOpen
  \bibfield  {author} {\bibinfo {author} {\bibfnamefont {W.~F.}\ \bibnamefont
  {Paxton}}, \bibinfo {author} {\bibfnamefont {A.}~\bibnamefont {Sen}}, \ and\
  \bibinfo {author} {\bibfnamefont {T.~E.}\ \bibnamefont {Mallouk}},\ }\href
  {\doibase 10.1002/chem.200500167} {\bibfield  {journal} {\bibinfo  {journal}
  {Chem. Eur. J.}\ }\textbf {\bibinfo {volume} {11}},\
  \bibinfo {pages} {6462} (\bibinfo {year} {2005})}\BibitemShut {NoStop}%
\bibitem [{\citenamefont {Mano}\ and\ \citenamefont {Heller}(2005)}]{Mano2005}%
  \BibitemOpen
  \bibfield  {author} {\bibinfo {author} {\bibfnamefont {N.}~\bibnamefont
  {Mano}}\ and\ \bibinfo {author} {\bibfnamefont {A.}~\bibnamefont {Heller}},\
  }\href {\doibase 10.1021/ja053937e} {\bibfield  {journal} {\bibinfo
  {journal} {J. Am. Chem. Soc.}\ }\textbf {\bibinfo
  {volume} {127}},\ \bibinfo {pages} {11574} (\bibinfo {year}
  {2005})}\BibitemShut {NoStop}%
\bibitem [{\citenamefont {Gibbs}\ and\ \citenamefont {Zhao}(2009)}]{Gibbs2009}%
  \BibitemOpen
  \bibfield  {author} {\bibinfo {author} {\bibfnamefont {J.~G.}\ \bibnamefont
  {Gibbs}}\ and\ \bibinfo {author} {\bibfnamefont {Y.-P.}\ \bibnamefont
  {Zhao}},\ }\href {\doibase 10.1063/1.3122346} {\bibfield  {journal} {\bibinfo
   {journal} {Appl. Phys. Lett.}\ }\textbf {\bibinfo {volume} {94}},\
  \bibinfo {pages} {163104} (\bibinfo {year} {2009})}\BibitemShut {NoStop}%
\bibitem [{\citenamefont {Jiang}\ \emph {et~al.}(2010)\citenamefont {Jiang},
  \citenamefont {Yoshinaga},\ and\ \citenamefont {Sano}}]{Jiang2010}%
  \BibitemOpen
  \bibfield  {author} {\bibinfo {author} {\bibfnamefont {H.-R.}\ \bibnamefont
  {Jiang}}, \bibinfo {author} {\bibfnamefont {N.}~\bibnamefont {Yoshinaga}}, \
  and\ \bibinfo {author} {\bibfnamefont {M.}~\bibnamefont {Sano}},\ }\href
  {\doibase 10.1103/PhysRevLett.105.268302} {\bibfield  {journal} {\bibinfo
  {journal} {Phys. Rev. Lett.}\ }\textbf {\bibinfo {volume} {105}},\
  \bibinfo {pages} {268302} (\bibinfo {year} {2010})}\BibitemShut {NoStop}%
\bibitem [{\citenamefont {Volpe}\ \emph {et~al.}(2011)\citenamefont {Volpe},
  \citenamefont {Buttinoni}, \citenamefont {Vogt}, \citenamefont
  {K\"{u}mmerer},\ and\ \citenamefont {Bechinger}}]{Volpe2011}%
  \BibitemOpen
  \bibfield  {author} {\bibinfo {author} {\bibfnamefont {G.}~\bibnamefont
  {Volpe}}, \bibinfo {author} {\bibfnamefont {I.}~\bibnamefont {Buttinoni}},
  \bibinfo {author} {\bibfnamefont {D.}~\bibnamefont {Vogt}}, \bibinfo {author}
  {\bibfnamefont {H.-J.}\ \bibnamefont {K\"{u}mmerer}}, \ and\ \bibinfo
  {author} {\bibfnamefont {C.}~\bibnamefont {Bechinger}},\ }\href {\doibase
  10.1039/c1sm05960b} {\bibfield  {journal} {\bibinfo  {journal} {Soft Matter}\
  }\textbf {\bibinfo {volume} {7}},\ \bibinfo {pages} {8810} (\bibinfo {year}
  {2011})}\BibitemShut {NoStop}%
\bibitem [{\citenamefont {Groot}\ and\ \citenamefont
  {Warren}(1997)}]{Groot1997}%
  \BibitemOpen
  \bibfield  {author} {\bibinfo {author} {\bibfnamefont {R.~D.}\ \bibnamefont
  {Groot}}\ and\ \bibinfo {author} {\bibfnamefont {P.~B.}\ \bibnamefont
  {Warren}},\ }\href {\doibase 10.1063/1.474784} {\bibfield  {journal}
  {\bibinfo  {journal} {J. Chem. Phys.}\ }\textbf {\bibinfo
  {volume} {107}},\ \bibinfo {pages} {4423} (\bibinfo {year}
  {1997})}\BibitemShut {NoStop}%
\bibitem [{\citenamefont {Levine}\ \emph {et~al.}(2000)\citenamefont {Levine},
  \citenamefont {Rappel},\ and\ \citenamefont {Cohen}}]{Levine2000}%
  \BibitemOpen
  \bibfield  {author} {\bibinfo {author} {\bibfnamefont {H.}~\bibnamefont
  {Levine}}, \bibinfo {author} {\bibfnamefont {W.-J.}\ \bibnamefont {Rappel}},
  \ and\ \bibinfo {author} {\bibfnamefont {I.}~\bibnamefont {Cohen}},\ }\href
  {\doibase 10.1103/PhysRevE.63.017101} {\bibfield  {journal} {\bibinfo
  {journal} {Phys. Rev. E}\ }\textbf {\bibinfo {volume} {63}},\ \bibinfo
  {pages} {017101} (\bibinfo {year} {2000})}\BibitemShut {NoStop}%
\bibitem [{\citenamefont {D'Orsogna}\ \emph {et~al.}(2006)\citenamefont
  {D'Orsogna}, \citenamefont {Chuang}, \citenamefont {Bertozzi},\ and\
  \citenamefont {Chayes}}]{DOrsogna2006}%
  \BibitemOpen
  \bibfield  {author} {\bibinfo {author} {\bibfnamefont {M.~R.}~\bibnamefont
  {D'Orsogna}}, \bibinfo {author} {\bibfnamefont {Y. L.}~\bibnamefont {Chuang}},
  \bibinfo {author} {\bibfnamefont {A.~L.}~\bibnamefont {Bertozzi}}, \ and\
  \bibinfo {author} {\bibfnamefont {L.~S.}~\bibnamefont {Chayes}},\ }\href
  {\doibase 10.1103/PhysRevLett.96.104302} {\bibfield  {journal} {\bibinfo
  {journal} {Phys. Rev. Lett.}\ }\textbf {\bibinfo {volume} {96}},\
  \bibinfo {pages} {104302} (\bibinfo {year} {2006})}\BibitemShut {NoStop}%
\bibitem [{\citenamefont {Chuang}\ \emph
  {et~al.}(2007{\natexlab{a}})\citenamefont {Chuang}, \citenamefont
  {D'Orsogna}, \citenamefont {Marthaler}, \citenamefont {Bertozzi},\ and\
  \citenamefont {Chayes}}]{Chuang2007}%
  \BibitemOpen
  \bibfield  {author} {\bibinfo {author} {\bibfnamefont {Y.-L.}\ \bibnamefont
  {Chuang}}, \bibinfo {author} {\bibfnamefont {M.~R.}\ \bibnamefont
  {D'Orsogna}}, \bibinfo {author} {\bibfnamefont {D.}~\bibnamefont
  {Marthaler}}, \bibinfo {author} {\bibfnamefont {A.~L.}\ \bibnamefont
  {Bertozzi}}, \ and\ \bibinfo {author} {\bibfnamefont {L.~S.}\ \bibnamefont
  {Chayes}},\ }\href {\doibase 10.1016/j.physd.2007.05.007} {\bibfield
  {journal} {\bibinfo  {journal} {Phys. D}\ }\textbf
  {\bibinfo {volume} {232}},\ \bibinfo {pages} {33} (\bibinfo {year}
  {2007}{\natexlab{a}})}\BibitemShut {NoStop}%
\bibitem [{\citenamefont {Chuang}\ \emph
  {et~al.}(2007{\natexlab{b}})\citenamefont {Chuang}, \citenamefont {Huang},
  \citenamefont {D'Orsogna},\ and\ \citenamefont {Bertozzi}}]{Chuang2007a}%
  \BibitemOpen
  \bibfield  {author} {\bibinfo {author} {\bibfnamefont {Y.-L.}\ \bibnamefont
  {Chuang}}, \bibinfo {author} {\bibfnamefont {Y.~R.}\ \bibnamefont {Huang}},
  \bibinfo {author} {\bibfnamefont {M.~R.}\ \bibnamefont {D'Orsogna}}, \ and\
  \bibinfo {author} {\bibfnamefont {A.~L.}\ \bibnamefont {Bertozzi}},\ }in\
  \href@noop {} {\emph {\bibinfo {booktitle} {IEEE International Conference on
  Robotics and Automation, Roma, Italy, 10--14 April}}},\ \bibinfo {series
  and number}  (\bibinfo {year} {2007})\BibitemShut
  {NoStop}%
\bibitem [{\citenamefont {Carrillo}\ \emph {et~al.}(2009)\citenamefont
  {Carrillo}, \citenamefont {D'Orsogna},\ and\ \citenamefont
  {Panferov}}]{Carrillo2009}%
  \BibitemOpen
  \bibfield  {author} {\bibinfo {author} {\bibfnamefont {J.}~\bibnamefont
  {Carrillo}}, \bibinfo {author} {\bibfnamefont {M.}~\bibnamefont {D'Orsogna}},
  \ and\ \bibinfo {author} {\bibfnamefont {V.}~\bibnamefont {Panferov}},\
  }\href {\doibase 10.3934/krm.2009.2.363} {\bibfield  {journal} {\bibinfo
  {journal} {Kinet. Relat. Models}\ }\textbf {\bibinfo {volume} {2}},\
  \bibinfo {pages} {363} (\bibinfo {year} {2009})}\BibitemShut {NoStop}%
\bibitem [{\citenamefont {Carrillo}\ \emph {et~al.}(2010)\citenamefont
  {Carrillo}, \citenamefont {Fornasier}, \citenamefont {Toscani},\ and\
  \citenamefont {Vecil}}]{Carrillo2010}%
  \BibitemOpen
  \bibfield  {author} {\bibinfo {author} {\bibfnamefont {J.~A.}\ \bibnamefont
  {Carrillo}}, \bibinfo {author} {\bibfnamefont {M.}~\bibnamefont {Fornasier}},
  \bibinfo {author} {\bibfnamefont {G.}~\bibnamefont {Toscani}}, \ and\
  \bibinfo {author} {\bibfnamefont {F.}~\bibnamefont {Vecil}},\ }in\ \href
  {\doibase 10.1007/978-0-8176-4946-3} {\emph {\bibinfo {booktitle}
  {Mathematical Modeling of Collective Behavior in Socio-Economic and Life
  Sciences, Modeling and Simulation in Science, Engineering and Technology}}},\
  \bibinfo {editor} {edited by\ \bibinfo {editor} {\bibfnamefont
  {G.}~\bibnamefont {Naldi}}, \bibinfo {editor} {\bibfnamefont
  {L.}~\bibnamefont {Pareschi}}, \ and\ \bibinfo {editor} {\bibfnamefont
  {G.}~\bibnamefont {Toscani}}}\ (\bibinfo  {publisher} {Birkh\"{a}user},\
  \bibinfo {address} {Boston},\ \bibinfo {year} {2010})\ pp.\ \bibinfo {pages}
  {297--336}\BibitemShut {NoStop}%
\bibitem [{\citenamefont {Plimpton}(1995)}]{Plimpton1995}%
  \BibitemOpen
  \bibfield  {author} {\bibinfo {author} {\bibfnamefont {S.}~\bibnamefont
  {Plimpton}},\ }\href@noop {} {\bibfield  {journal} {\bibinfo  {journal}
  {J. Comput. Phys.}\ }\textbf {\bibinfo {volume} {117}},\
  \bibinfo {pages} {1} (\bibinfo {year} {1995})}\BibitemShut {NoStop}%
\bibitem [{\citenamefont {{LAMMPS (Large-scale Atomic/Molecular Massively
  Parallel Simulator)}}(2013)}]{LAMMPS2013}%
  \BibitemOpen
  \bibfield  {author} {\bibinfo {author} {\bibnamefont {{LAMMPS (Large-scale
  Atomic/Molecular Massively Parallel Simulator)}}},\ }\href@noop {} {\enquote
  {\bibinfo {title} {http://lammps.sandia.gov},}\ } (\bibinfo {year}
  {2013})\BibitemShut {NoStop}%
\bibitem [{\citenamefont {Verlet}(1967)}]{Verlet1967}%
  \BibitemOpen
  \bibfield  {author} {\bibinfo {author} {\bibfnamefont {L.}~\bibnamefont
  {Verlet}},\ }\href@noop {} {\bibfield  {journal} {\bibinfo  {journal}
  {Phys. Rev.}\ }\textbf {\bibinfo {volume} {159}},\ \bibinfo {pages} {98}
  (\bibinfo {year} {1967})}\BibitemShut {NoStop}%
%
%
\bibitem [{\citenamefont {Woodhouse}\ and\ \citenamefont
  {Goldstein}(2012)}]{Woodhouse2012}%
  \BibitemOpen
  \bibfield  {author} {\bibinfo {author} {\bibfnamefont {F.~G.}~\bibnamefont
  {Woodhouse}}\ and\ \bibinfo {author} {\bibfnamefont {R.~E.}~\bibnamefont
  {Goldstein}},\ }\href@noop {} {\bibfield  {journal} {\bibinfo  {journal}
  {Phys. Rev. Lett.}\ }\textbf {\bibinfo {volume} {109}},\ \bibinfo {pages} {168105}
  (\bibinfo {year} {2012})}\BibitemShut {NoStop}%
\bibitem [{\citenamefont {Wioland}\ \emph {et~al.}(2013)\citenamefont
  {Wioland}, \citenamefont {Woodhouse}, \citenamefont {Dunkel}, \citenamefont {Kessler},\ and\
  \citenamefont {Goldstein}}]{Wioland2013a}%
  \BibitemOpen
  \bibfield  {author} {\bibinfo {author} {\bibfnamefont {H.}~\bibnamefont
  {Wioland}}, \bibinfo {author} {\bibfnamefont {F.~G.}~\bibnamefont {Woodhouse}},
  \bibinfo {author} {\bibfnamefont {J.}~\bibnamefont {Dunkel}}, {\bibfnamefont {J.~O.}~\bibnamefont {Kessler}}, \ and\ \bibinfo {author} {\bibfnamefont {R.~E.}~\bibnamefont {Goldstein}},\ }\href
  {\doibase 10.1103/PhysRevLett.110.268102} {\bibfield  {journal} {\bibinfo
  {journal} {Phys. Rev. Lett.}\ }\textbf {\bibinfo {volume} {110}},\
  \bibinfo {pages} {268102} (\bibinfo {year} {2013})}\BibitemShut {NoStop}%
%
%
%
%
\bibitem [{\citenamefont {Metzler}\ and\ \citenamefont
  {Klafter}(2000)}]{Metzler2000}%
  \BibitemOpen
  \bibfield  {author} {\bibinfo {author} {\bibfnamefont {R.}~\bibnamefont
  {Metzler}}\ and\ \bibinfo {author} {\bibfnamefont {J.}~\bibnamefont
  {Klafter}},\ }\href@noop {} {\bibfield  {journal} {\bibinfo  {journal}
  {Phys. Rep.}\ }\textbf {\bibinfo {volume} {339}},\ \bibinfo {pages} {1}
  (\bibinfo {year} {2000})}\BibitemShut {NoStop}%
\bibitem [{\citenamefont {Einstein}(1905)}]{Einstein1905}%
  \BibitemOpen
  \bibfield  {author} {\bibinfo {author} {\bibfnamefont {A.}~\bibnamefont
  {Einstein}},\ }\href@noop {} {\bibfield  {journal} {\bibinfo  {journal}
  {Ann. Phys.}\ }\textbf {\bibinfo {volume} {17}},\ \bibinfo {pages}
  {549} (\bibinfo {year} {1905})}\BibitemShut {NoStop}%
\bibitem [{\citenamefont {Wu}\ and\ \citenamefont {Libchaber}(2000)}]{Wu2000}%
  \BibitemOpen
  \bibfield  {author} {\bibinfo {author} {\bibfnamefont {X.-L.}\ \bibnamefont
  {Wu}}\ and\ \bibinfo {author} {\bibfnamefont {A.}~\bibnamefont {Libchaber}},\
  }\href {http://www.ncbi.nlm.nih.gov/pubmed/11177879
  http://link.aps.org/doi/10.1103/PhysRevLett.84.3017} {\bibfield  {journal}
  {\bibinfo  {journal} {Phys. Rev. Lett.}\ }\textbf {\bibinfo {volume}
  {84}},\ \bibinfo {pages} {3017} (\bibinfo {year} {2000})}\BibitemShut
  {NoStop}%
\bibitem [{\citenamefont {Gr\'{e}goire}\ \emph {et~al.}(2001)\citenamefont
  {Gr\'{e}goire}, \citenamefont {Chat\'{e}},\ and\ \citenamefont
  {Tu}}]{Gregoire2001a}%
  \BibitemOpen
  \bibfield  {author} {\bibinfo {author} {\bibfnamefont {G.}~\bibnamefont
  {Gr\'{e}goire}}, \bibinfo {author} {\bibfnamefont {H.}~\bibnamefont
  {Chat\'{e}}}, \ and\ \bibinfo {author} {\bibfnamefont {Y.}~\bibnamefont
  {Tu}},\ }\href {\doibase 10.1103/PhysRevLett.86.556} {\bibfield  {journal}
  {\bibinfo  {journal} {Phys. Rev. Lett.}\ }\textbf {\bibinfo {volume}
  {86}},\ \bibinfo {pages} {556} (\bibinfo {year} {2001})}\BibitemShut
  {NoStop}%
\bibitem [{\citenamefont {Wu}\ and\ \citenamefont {Libchaber}(2001)}]{Wu2001a}%
  \BibitemOpen
  \bibfield  {author} {\bibinfo {author} {\bibfnamefont {X.-L.}\ \bibnamefont
  {Wu}}\ and\ \bibinfo {author} {\bibfnamefont {A.}~\bibnamefont {Libchaber}},\
  }\href {\doibase 10.1103/PhysRevLett.86.557} {\bibfield  {journal} {\bibinfo
  {journal} {Phys. Rev. Lett.}\ }\textbf {\bibinfo {volume} {86}},\
  \bibinfo {pages} {557} (\bibinfo {year} {2001})}\BibitemShut {NoStop}%
  %
  %
  \bibitem [{\citenamefont {Tao}\ and\ \citenamefont {Kapral}(2009)}]{Tao2009}%
  \BibitemOpen
  \bibfield  {author} {\bibinfo {author} {\bibfnamefont {Y.-G.}\ \bibnamefont
  {Tao}}\ and\ \bibinfo {author} {\bibfnamefont {R.}~\bibnamefont {Kapral}},\
  }\href {\doibase 10.1002/cphc.200800829} {\bibfield  {journal} {\bibinfo
  {journal} {ChemPhysChem}\ }\textbf {\bibinfo {volume} {10}},\ \bibinfo
  {pages} {770} (\bibinfo {year} {2009})}\BibitemShut {NoStop}%
\bibitem [{\citenamefont {Kapral}(2013)}]{Kapral2013}%
  \BibitemOpen
  \bibfield  {author} {\bibinfo {author} {\bibfnamefont {R.}~\bibnamefont
  {Kapral}},\ }\href {\doibase 10.1063/1.4773981} {\bibfield  {journal}
  {\bibinfo  {journal} {J.\ Chem.\ Phys.}\ }\textbf {\bibinfo
  {volume} {138}},\ \bibinfo {pages} {020901} (\bibinfo {year}
  {2013})}\BibitemShut {NoStop}%
\bibitem [{\citenamefont {Wu}\ \emph {et~al.}(2013)\citenamefont {Wu},
  \citenamefont {Wu}, \citenamefont {He}, \citenamefont {Lin}, \citenamefont
  {Sun},\ and\ \citenamefont {He}}]{Wu2013}%
  \BibitemOpen
  \bibfield  {author} {\bibinfo {author} {\bibfnamefont {Z.}~\bibnamefont
  {Wu}}, \bibinfo {author} {\bibfnamefont {Y.}~\bibnamefont {Wu}}, \bibinfo
  {author} {\bibfnamefont {W.}~\bibnamefont {He}}, \bibinfo {author}
  {\bibfnamefont {X.}~\bibnamefont {Lin}}, \bibinfo {author} {\bibfnamefont
  {J.}~\bibnamefont {Sun}}, \ and\ \bibinfo {author} {\bibfnamefont
  {Q.}~\bibnamefont {He}},\ }\href {\doibase 10.1002/anie.201301643} {\bibfield
   {journal} {\bibinfo  {journal} {Angew.\ Chem.}\ }\textbf {\bibinfo
  {volume} {52}},\ \bibinfo {pages} {7000} (\bibinfo {year}
  {2013})}\BibitemShut {NoStop}%
\bibitem [{\citenamefont {Giomi}\ \emph {et~al.}(2013)\citenamefont {Giomi},
  \citenamefont {Hawley-Weld},\ and\ \citenamefont {Mahadevan}}]{Giomi2013}%
  \BibitemOpen
  \bibfield  {author} {\bibinfo {author} {\bibfnamefont {L.}~\bibnamefont
  {Giomi}}, \bibinfo {author} {\bibfnamefont {N.}~\bibnamefont {Hawley-Weld}},
  \ and\ \bibinfo {author} {\bibfnamefont {L.}~\bibnamefont {Mahadevan}},\
  }\href@noop {} {\bibfield  {journal} {\bibinfo  {journal} {Proc.\ 
  R.\ Soc.\ A}\ }\textbf {\bibinfo {volume} {469}},\ \bibinfo {pages}
  {20120637} (\bibinfo {year} {2013})}\BibitemShut {NoStop}%
\bibitem [{\citenamefont {Li}\ and\ \citenamefont {Zhang}(2013)}]{Li2013}%
  \BibitemOpen
  \bibfield  {author} {\bibinfo {author} {\bibfnamefont {H.}~\bibnamefont
  {Li}}\ and\ \bibinfo {author} {\bibfnamefont {H.~P.}\ \bibnamefont {Zhang}},\
  }\href {\doibase 10.1209/0295-5075/102/50007} {\bibfield  {journal} {\bibinfo
   {journal} {Europhys.\ Lett.}\ }\textbf {\bibinfo {volume} {102}},\ \bibinfo {pages}
  {50007} (\bibinfo {year} {2013})}\BibitemShut {NoStop}%
%
  %
  %
\bibitem [{\citenamefont {Ghosh}\ \emph {et~al.}(2013)\citenamefont
  {Ghosh}, \citenamefont {Misko}, \citenamefont {Marchesoni},\ and\
  \citenamefont {Nori}}]{Ghosh2013}%
  \BibitemOpen
  \bibfield  {author} {\bibinfo {author} {\bibfnamefont {P.~K.}~\bibnamefont
  {Ghosh}}, \bibinfo {author} {\bibfnamefont {V.~R.}~\bibnamefont {Misko}},
  {\bibfnamefont {F.}~\bibnamefont {Marchesoni}}, \ and\ \bibinfo {author} {\bibfnamefont {F.}~\bibnamefont {Nori}},\ }\href
  {\doibase 10.1103/PhysRevLett.110.268301} {\bibfield  {journal} {\bibinfo
  {journal} {Phys. Rev. Lett.}\ }\textbf {\bibinfo {volume} {110}},\
  \bibinfo {pages} {268301} (\bibinfo {year} {2013})}\BibitemShut {NoStop}%
\bibitem [{\citenamefont {Patra}\ \emph {et~al.}(2013)\citenamefont {Patra},
  \citenamefont {Sengupta}, \citenamefont {Duan}, \citenamefont {Zhang},
  \citenamefont {Pavlick},\ and\ \citenamefont {Sen}}]{Patra2013}%
  \BibitemOpen
  \bibfield  {author} {\bibinfo {author} {\bibfnamefont {D.}~\bibnamefont
  {Patra}}, \bibinfo {author} {\bibfnamefont {S.}~\bibnamefont {Sengupta}},
  \bibinfo {author} {\bibfnamefont {W.}~\bibnamefont {Duan}}, \bibinfo {author}
  {\bibfnamefont {H.}~\bibnamefont {Zhang}}, \bibinfo {author} {\bibfnamefont
  {R.}~\bibnamefont {Pavlick}}, \ and\ \bibinfo {author} {\bibfnamefont
  {A.}~\bibnamefont {Sen}},\ }\href {\doibase 10.1039/c2nr32600k} {\bibfield
  {journal} {\bibinfo  {journal} {Nanoscale}\ }\textbf {\bibinfo {volume}
  {5}},\ \bibinfo {pages} {1273} (\bibinfo {year} {2013})}\BibitemShut
  {NoStop}%
\bibitem [{\citenamefont {Soler}\ \emph {et~al.}(2013)\citenamefont {Soler},
  \citenamefont {Magdanz}, \citenamefont {Fomin}, \citenamefont {Sanchez},\
  and\ \citenamefont {Schmidt}}]{Soler2013a}%
  \BibitemOpen
  \bibfield  {author} {\bibinfo {author} {\bibfnamefont {L.}~\bibnamefont
  {Soler}}, \bibinfo {author} {\bibfnamefont {V.}~\bibnamefont {Magdanz}},
  \bibinfo {author} {\bibfnamefont {V.~M.}\ \bibnamefont {Fomin}}, \bibinfo
  {author} {\bibfnamefont {S.}~\bibnamefont {Sanchez}}, \ and\ \bibinfo
  {author} {\bibfnamefont {O.~G.}\ \bibnamefont {Schmidt}},\ }\href
  {http://pubs.acs.org/doi/abs/10.1021/nn405075d} {\bibfield  {journal}
  {\bibinfo  {journal} {ACS Nano}\ }\textbf {\bibinfo {volume} {7}},\ \bibinfo
  {pages} {9611} (\bibinfo {year} {2013})}\BibitemShut {NoStop}%
\end{thebibliography}
%


\end{document}